\documentclass[preprint,12pt]{elsarticle}

\usepackage{amssymb}
\usepackage{amsthm}

\usepackage{float}
\usepackage{color}
\usepackage{amsmath}
\usepackage{graphicx, subfigure}
\usepackage{pslatex}
\usepackage{relsize}
\usepackage{bm}
\usepackage{verbatim}

\usepackage{booktabs}
\usepackage{xcolor}
\usepackage{soul}

\usepackage[utf8]{inputenc}
\usepackage[T1]{fontenc}

\biboptions{numbers,sort&compress}

\begin{document}

\begin{frontmatter}



\title{Automated calculations of exchange magnetostriction}


\author[a]{P.~Nieves\corref{author}}
\author[a]{S.~Arapan}
\author[b,c]{S.~H.~Zhang}
\author[a]{A.~P.~K\k{a}dzielawa}
\author[b,c]{R.~F.~Zhang }
\author[a]{D.~Legut}

\cortext[author] {Corresponding author.\\\textit{E-mail address:} pablo.nieves.cordones@vsb.cz}
\address[a]{IT4Innovations, V\v{S}B - Technical University of Ostrava, 17. listopadu 2172/15, 70800 Ostrava-Poruba, Czech Republic}
\address[b]{School of Materials Science and Engineering, Beihang University, Beijing 100191, PR China}
\address[c]{Center for Integrated Computational Materials Engineering, International Research Institute for Multidisciplinary Science, Beihang University, Beijing
100191, PR China}

\begin{abstract}

We present a methodology based on deformations of the unit cell that allows to compute the isotropic magnetoelastic constants, isotropic magnetostrictive coefficients and spontaneous volume magnetostriction associated to the exchange magnetostriction. This method is implemented in the python package MAELAS (v3.0), so that it can be used to obtain these quantities by first-principles calculations and classical spin-lattice models in an automated way. We show that the required reference state to obtain the spontaneous volume magnetostriction combines the equilibrium volume of the paramagnetic state and magnetic order of the ground state. We identify an error in the theoretical expression of the isotropic magnetostrictive coefficient $\lambda^{\alpha1,0}$ for uniaxial crystals given in previous publications, which is corrected in this work. The presented computational tool may be helpful to provide a better understanding and characterization of the relationship between the exchange interaction and magnetoelasticity.

\end{abstract}


\end{frontmatter}


\section{Introduction}
\label{section:Intro}

The isotropic magnetoelastic constants ($b^{iso}$) give a contribution to the magnetoelastic energy that depends on the strain but not on the magnetization direction at saturated state. They are mainly originated by the volume dependence of both atomic magnetic moments and exchange interaction \cite{booktremolet}, so that are responsible for the so-called exchange magnetostriction\cite{WASSERMAN1990237,ANDREEV199559,booktremolet}. Other less significant contributions to $b^{iso}$ are the isotropic contribution of the dipolar magnetostriction (form effect) and crystal effect \cite{booktremolet}. From the minimization of the elastic energy and isotropic contribution of magnetoelastic energy with respect to strain, one can derive the isotropic contribution to the fractional change in length ($[l-l_0]/l_0$) which is characterized by the isotropic magnetostrictive coefficients $\lambda^{iso}$. These quantities lead to the isotropic contribution to the spontaneous volume magnetostriction ($\omega_s$)\cite{WASSERMAN1990237,ANDREEV199559,booktremolet}, defined as the fractional volume change between magnetically ordered and paramagnetic (PM) state ($\omega_s=[V_{FM}-V_{PM}]/V_{PM}$) \cite{khmelevskyi}. This property is responsible for interesting features like anomalies in the thermal expansion coefficient in magnetic materials\cite{WASSERMAN1990237,ANDREEV199559}. For example, Invar alloy (Fe$_{65}$Ni$_{35}$) exhibits a very large $\omega_s$ ($\sim 10^{-2}$) that cancels the normal thermal expansion, leading to nearly zero net thermal expansion over a broad range of temperatures \cite{WASSERMAN1990237}. Nowadays, Invar alloys and their extensions (Super-Invar, Stainless-Invar, Elinvar, Super-Elinvar, Co-Elinvar)\cite{WASSERMAN1990237} are widely used in many  commercial applications such as 
precision machine tools, precision pendulums, precision capacitors, precision moulds, transistor
bases, lead frames for integrated circuits, thermostats, bending meters, gravity meters, flow meters, astronomical telescopes, seismographic devices, microwave guides, resonant cavities, laser light sources and radar echo boxes \cite{WASSERMAN1990237}.  Exchange magnetostriction might also play an important role in novel magnetic phenomena like laser-induced ultrafast magnetism \cite{Reid2018}.

At an arbitrary temperature $T$, we have $\omega_s(T)=(V_{FM}(T)-V_{PM}(T))/V_{PM}(T)$. Above the Curie or N\'{e}el temperature ($T>T_c$), the spontaneous exchange magnetostriction is zero $\omega_s(T)=0$. Below the Curie or N\'{e}el temperature ($T<T_c$), one needs to know the equilibrium volume at hypothetical PM-like state $V_{PM}(T)$ at $T<T_c$, which is difficult to characterize experimentally and theoretically \cite{ANDREEV199559,WASSERMAN1990237,khmelevskyi}. In experiment, such equilibrium volume is typically estimated by an extrapolation using the Debye theory and the
Gr\"{u}neisen relation from above the Curie or N\'{e}el temperature\cite{ANDREEV199559,WASSERMAN1990237}. On the other hand, the PM state at zero-temperature can be studied with first-principles calculations through the Stoner model\cite{kake1986,shimizu1978}, disordered local moment (DLM) approach\cite{khmelevskyi,Alling,khem2004,TUREK2005357} and special-quasirandom structures (SQS)\cite{Ikeda}. For example, first principles calculations of $\omega_s$ showed that  the Stoner model might overestimate  it in body-centered cubic (bcc) Fe \cite{kake1986,khmelevskyi}, while a better quantitative agreement with experiment is achieved through the disordered local moment (DLM) approach\cite{khmelevskyi}.

In practice, the theoretical study of the PM state at zero-temperature and $\omega_s$ with current available methods is possible but is not easy. So in this sense, it would be desirable to find alternative ways to compute these quantities which could be easily  automatized\cite{AELAS,maelas_publication2021,maelas_v2},  making it more suitable for high-throughput screening. For example, such kind of strategy might be helpful to further improve Invar alloys or to discover new families of magnetic materials with very large $\omega_s$ ($>10^{-2}$).  In previous versions of the python package   MAELAS\cite{maelas_publication2021,maelas_v2}, it was possible to compute anisotropic magnetoelastic constants ($b^{ani}$) and magnetostrictive coefficients ($\lambda^{ani}$) in automated way but not the isotropic ones. Here, we release a new version of MAELAS (v3.0) where we extend its capabilities by implementing a methodology to compute $b^{iso}$, $\lambda^{iso}$ and $\omega_s$. To do so, we make use of the universal notation proposed by E. du Tremolet de Lacheisserie \cite{booktremolet} which, thanks to its rigorous theoretical derivation based on the framework of group theory, naturally decomposes the definition of the magnetoelastic constants arising from isotropic and anisotropic magnetic interactions. The source files of this new version are available in GitHub repository \cite{Maelas}.

The paper is organized as follows. In Section \ref{section:method}, we describe the methodology to compute $b^{iso}$, $\lambda^{iso}$ and $\omega_s$, while Section \ref{section:comp_details} is devoted to technical details about its implementation in MAELAS. This method is benchmarked in Section \ref{section:test}. In Section \ref{section:limit}, we discuss about some limitations of this version and future perspectives. The paper ends with a summary of the main conclusions and future perspectives (Section \ref{section:con}).

\section{Methodology}
\label{section:method}

We consider two cases depending on whether the anisotropic magnetic interactions are included or not. 

\subsection{Including only isotropic magnetic interactions}
\label{subsection:method_iso}

In magnetic materials where the main source of magnetic anisotropy arises from spin-orbit coupling (SOC), one can easily compute the total energy without anisotropic magnetic interactions through first-principles calculations by just switching off the SOC. In this case $b^{ani}=0$ and the magnetoelastic energy contains only the isotropic term. This approach also makes it possible to speed up this type of task, since including SOC is more computationally demanding.

The method to compute $b^{iso}$, $\lambda^{iso}$ and $\omega_s$, including only isotropic magnetic interactions, is derived from the total spin-polarized energy ($E$) including elastic ($E_{el}$) and isotropic magnetoelastic ($E^{iso}_{me}$), that is,
\begin{equation}
    E(\boldsymbol{\epsilon})=E_{el}(\boldsymbol{\epsilon})+E^{iso}_{me}(\boldsymbol{\epsilon}),
\label{eq:E_tot}
\end{equation}
where $\boldsymbol{\epsilon}$ is the strain tensor. Here, the elastic energy is considered up to second order in the strain, while the isotropic magnetoelastic energy contains only linear terms in the strain and no dependence on magnetization direction $\boldsymbol{\alpha}$. This means that we assume that the total energy of the system is invariant under rotations of the magnetization at the saturated state. Hence, the anisotropic magnetoelastic must be negligible ($E^{ani}_{me}\ll E^{iso}_{me}$).  For example, this assumption could be accomplished by not including the SOC in the calculation of the energy for magnetic materials where the main source of magnetic anisotropy arises from SOC. Note that $E^{iso}_{me}$ must be purely originated by isotropic magnetic interactions like isotropic exchange. Unfortunately, the definition of magnetoelastic constants in some conventions mixes the contribution of both isotropic and anisotropic magnetic interactions, so that it is not possible to write $E^{iso}_{me}$ in terms of these magnetoelastic constants. This problem can be overcome by using the universal definition of the magnetoelastic constants proposed by E. du Tremolet de Lacheisserie on the basis of group theory\cite{booktremolet}, which fully decouples the contribution of isotropic and anisotropic magnetic interactions in the definition of the magnetoelastic constants. The explicit form of $E^{iso}_{me}$, based on Lacheisserie convention \cite{booktremolet}, for each supported crystal symmetry in MAELAS is shown in \ref{app_matrix}. In Section \ref{section:limit}, we discuss about possible limitations of the method presented here.

\begin{figure}[h!]
\centering
\includegraphics[width=\columnwidth ,angle=0]{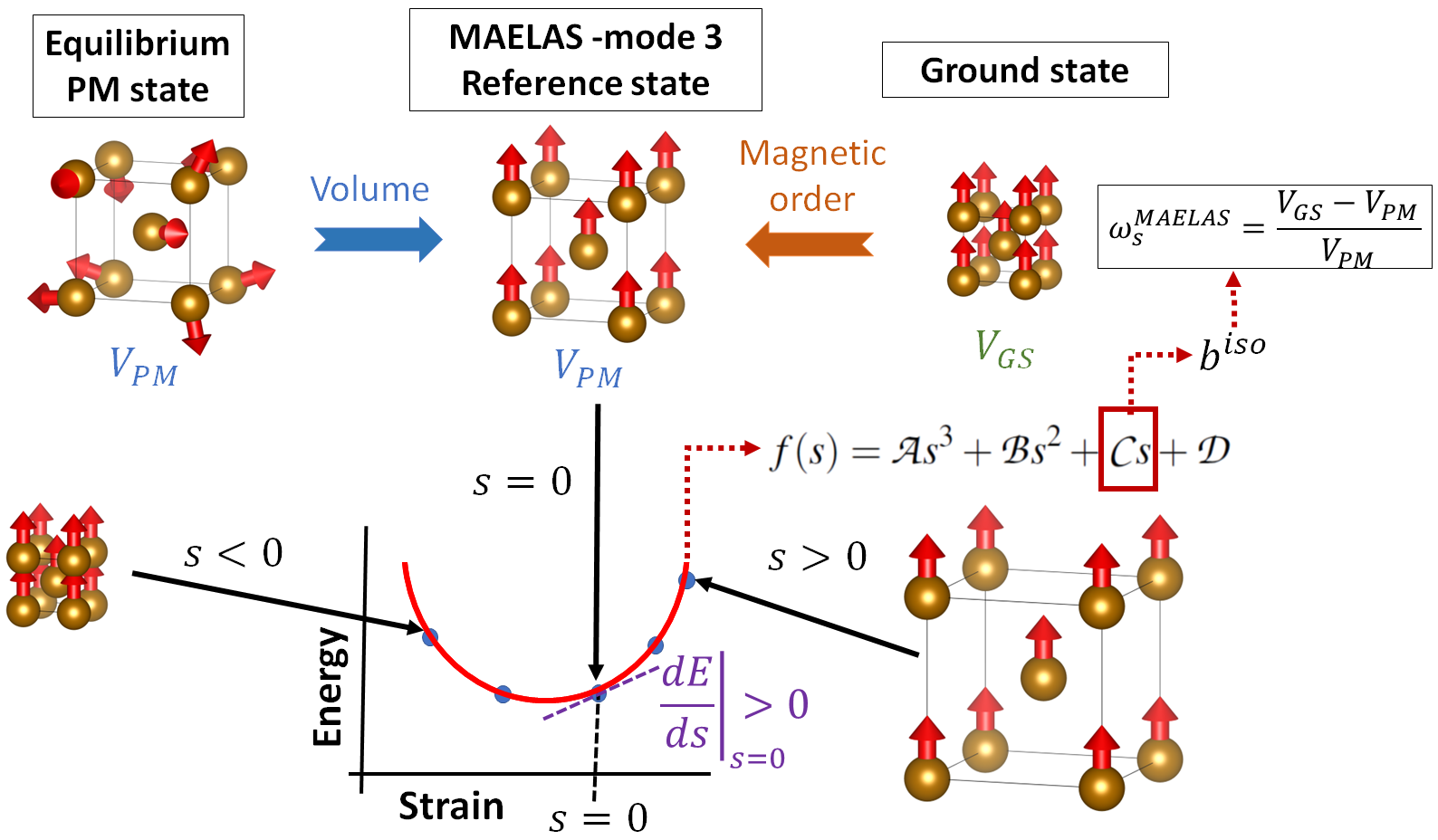}
\caption{Schematic diagram showing the reference state (unstrained unit cell $s=0$) required by the -mode 3 of MAELAS to generate the deformed unit cells, and derive the isotropic magnetoelastic constants $b^{iso}$. This reference state is constructed by combining the equilibrium lattice parameters of the PM state and the magnetic order of the GS. In this figure we assume that $\omega_s<0$ ($V_{GS}<V_{PM}$), and the GS of the material is FM ($V_{GS}=V_{FM}$). If the GS is AFM, then the FM configurations in these unit cells should be changed to AFM. }
\label{fig:ref_mode3}
\end{figure}

\begin{figure}[h!]
\centering
\includegraphics[width=\columnwidth ,angle=0]{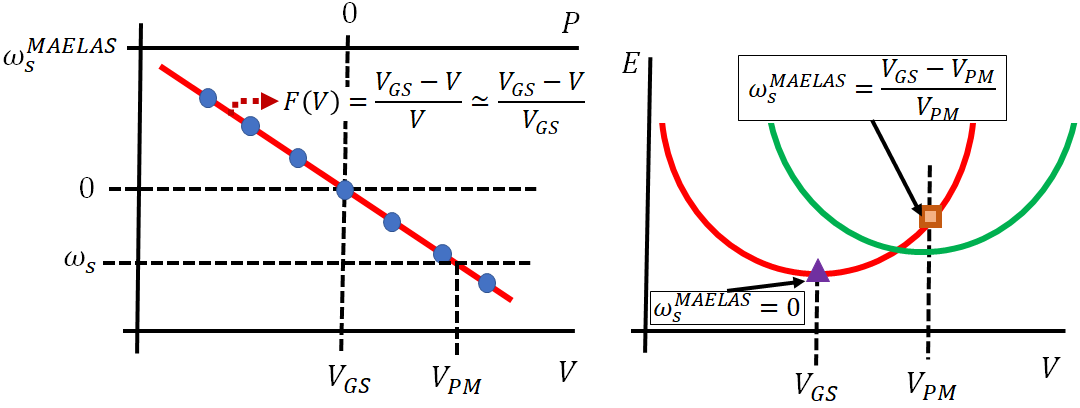}
\caption{(Left) Spontaneous volume magnetostriction obtained with MAELAS $\omega_s^{MAELAS}(V)$ versus the volume of the reference unit cell ($V$) from which MAELAS generates the deformations. (Right) Equation of the state using the magnetic order of the GS (red) and the PM order (green). Orange square shows the reference state required by -mode 3 of  MAELAS to give the spontaneous volume magnetostriction $\omega_s^{MAELAS}(V_{PM})=\omega_s=(V_{GS}-V_{PM})/V_{PM}$. Purple triangle corresponds to the GS. If the GS is used as reference state, then -mode 3 of MAELAS gives zero spontaneous volume magnetostriction $\omega_s^{MAELAS}(V_{GS})=0$ since the exchange magnetostriction is already released from the PM state.}
\label{fig:vpm_maelas}
\end{figure}

The basic idea of this new method is to compute the total spin-polarized energy (Eq.\ref{eq:E_tot}) for a set of deformed unit cells in such a way that we can get the i-th isotropic magnetoelastic constant $b_i^{iso}$  from a polynomial fitting of the energy versus strain data
\begin{equation}
   \frac{1}{V_0} E(\boldsymbol{\epsilon}^i(s))=\Phi_i(C_{ij})s^2+\Gamma_i b^{iso}_i s + \frac{1}{V_0} E_0,
\label{eq:dE_tot}
\end{equation}
where $V_0$ is the equilibrium volume of the reference unstrained state ($s=0$), $\Gamma_i$ is a real number, $\Phi_i$ depends on the elastic constants $C_{ij}$, $E_0$ is the energy at the unstrained state, and $s$ is the parameter used to parameterize the strain tensor $\boldsymbol{\epsilon}^i(s)$. In practice, Eq. \ref{eq:dE_tot} is fitted to a third order polynomial
\begin{equation}
    f(s)=\mathcal{A}s^3+\mathcal{B}s^2+\mathcal{C}s+\mathcal{D},
\label{eq:dE_tot_fit}
\end{equation}
where $\mathcal{A}$, $\mathcal{B}$, $\mathcal{C}$ and $\mathcal{D}$ are fitting parameters, so that the i-th isotropic magnetoelastic constant $b_i^{iso}$ is given by
\begin{equation}
    b_i^{iso}=\frac{\mathcal{C}}{\Gamma_i}.
\label{eq:b_i_sol}
\end{equation}
Additionally, we note that the fitting parameter $\mathcal{B}=\Phi_i(C_{ij})$ contains information about the elastic constants, which is the basis of the program AELAS to derive $C_{ij}$\cite{AELAS}. The computed elastic constants $C_{ij}$ with AELAS (energy-strain method) may include high order magnetic corrections from the second order in strain isotropic magnetoelastic constants\cite{Rouchy1979,booktremolet,nieves_sound2022}. In Section \ref{subsection:lmp}, we show that a third order polynomial fit (Eq.\ref{eq:dE_tot_fit}) is more accurate than second order to compute $b_i^{iso}$ and $\omega_s$ in a broad range of applied strains. In Table \ref{tab:method_data} we show the selected set of $b_i^{iso}$, parameterized strain tensor $\boldsymbol{\epsilon}^i(s)$ and the corresponding value of $\Gamma_i$ that fulfils Eq. \ref{eq:dE_tot}. If the elastic constants are also computed\cite{AELAS}, then one can also calculate the isotropic magnetostrictive coefficients ($\lambda^{iso}$) and isotropic contribution to spontaneous volume magnetostriction $\omega^{iso}_s$ using the theoretical equations derived from magnetoelasticity ($\lambda^{iso}(b^{iso}_{k},C_{ij})$ and $\omega^{iso}_s(\lambda^{iso})$), see \ref{app_matrix}. We implemented this method in the python package MAELAS for all supported crystal symmetries in previous version of MAELAS \cite{maelas_publication2021,maelas_v2}. This new method is executed in MAELAS by using tag -mode 3 in the command line, see more details in Section \ref{section:comp_details}. The calculated $b^{iso}$ by this approach depends strongly on the reference state (unstrained unit cell $s=0$) used to generate the deformed unit cells, so that it plays a very important role. To analyze this dependence, let us use the Taylor series of function $f(s)$ at $s=0$ (reference unstrained state). Doing so, we find that $b_i^{iso}$ may be also written as
\begin{equation}
    b_i^{iso}=\frac{1}{\Gamma_i}\frac{df(s)}{ds}\Bigg\vert_{s=0}=\frac{1}{V_0\Gamma_i}\frac{dE(\boldsymbol{\epsilon}^i(s))}{ds}\Bigg\vert_{s=0}.
\label{eq:taylor_b_i_sol}
\end{equation}
Here, we see that using the ground state (GS) as reference state (unstrained unit cell $s=0$) leads to a null exchange magnetostriction $b^{iso}=0$ since $dE/ds(s=0)=0$. In this case, any finite value of $b^{iso}$ given by MAELAS may be related to numerical effects or small deviation from the exact GS (state at zero external pressure) rather than a physical effect. Hence, it is important to clarify which reference state gives the correct $b_i^{iso}$ and $\omega^{iso}_s$ with respect to the PM state. To do so, let us consider a reference state with the same magnetic order as the GS, i.e. ferromagnetic (FM) or antiferromagnetic (AFM), and an arbitrary volume $V$ close to the equilibrium volume of the GS ($V_{GS}$), where atoms were previously relaxed to the equilibrium position at this constant volume $V$. Applying this methodology to this reference state, we find $b^{iso}_{MAELAS}(V)$ and $C_{ij}^{AELAS}(V)$, which gives the isotropic magnetostrictive coefficients ($\lambda^{iso}_{MAELAS}(V)$). These isotropic magnetostrictive coefficients describe the fractional change in length $(l_{GS}-l)/l$, where $l_{GS}$ and $l$ are the length of the material along a measuring direction $\boldsymbol{\beta}$ at the GS ($V_{GS}$) and reference state ($V$), respectively. From the fractional change in length, one can immediately derive the fractional change in volume using a geometrical analysis. For example, in cubic crystals we have $\Delta V/V=3\Delta l/l +O(\lambda^2)$. Consequently, the fractional change in volume given by this method using an arbitrary reference state ($V$) is
\begin{equation}
    \omega_s^{MAELAS}(V)=\frac{V_{GS}-V}{V},
\label{eq:ws_maelas_gs}
\end{equation}
where $V_{GS}=V_{FM}$ if the GS is FM, or $V_{GS}=V_{AFM}$ if the ground GS is AFM. We see that this method gives the fractional change in volume of the GS with respect to the volume of the reference state. Consequently, the real spontaneous volume magnetostriction ($\omega_s$) will be obtained using a reference state at the PM volume ($V_{PM}$)
\begin{equation}
    \omega_s^{MAELAS}(V_{PM})=\frac{V_{GS}-V_{PM}}{V_{PM}}\equiv \omega_s.
\label{eq:ws_maelas}
\end{equation}
In summary, the correct reference state should be constructed by combining the equilibrium volume of the PM state ($V_{PM}$) and the magnetic order of the GS (FM or AFM), see Fig.\ref{fig:ref_mode3}. The atoms at this reference state should be  relaxed to the equilibrium position at constant volume $V_{PM}$ keeping the same magnetic order as the GS before the generation of the deformations with MAELAS. The energy of deformed unit cells generated by MAELAS should be evaluated using the same magnetic order as the GS (FM or AFM). In the case of non-cubic crystals, one also needs that the reference state has the equilibrium lattice parameters of the PM state ($a_{PM}$, $b_{PM}$ and $c_{PM}$) in order to obtain the correct fractional change in length with respect to the PM state $(l_{GS}-l_{PM})/l_{PM}$ described by the isotropic magnetostrictive coefficients $\lambda^{iso}$. Hence, we see that the equilibrium lattice of the PM state must be known to build this reference state. In general, the first-principles calculation of the PM state is not straightforward with available theoretical methods. Unfortunately, this fact may limit the applicability of the presented method like in high-throughput screening approaches for exchange magnetostriction. The spontaneous volume magnetostriction computed by MAELAS follows approximately a linear behaviour with respect to the volume of the reference state ($V$) close to $V_{GS}$, which arises from a Taylor expansion of Eq.\ref{eq:ws_maelas_gs} around $V_{GS}$
\begin{equation}
    \omega_s^{MAELAS}(V)=\frac{V_{GS}-V}{V}\simeq\frac{V_{GS}-V}{V_{GS}}+O\left(\left[\frac{V_{GS}-V}{V_{GS}}\right]^2\right), \quad \frac{\vert V_{GS}-V\vert}{V_{GS}}\ll 1.
\label{eq:ws_maelas_gs_taylor}
\end{equation}
This dependence is schematically represented in Fig.\ref{fig:vpm_maelas}.

\begin{table}[]
\centering
\caption{Selected parameters in the new method implemented in MAELAS version 3.0 to calculate the isotropic magnetoelastic constants. The first column shows the crystal system using the notation of Wallace \cite{Wallace,mouhat} (I/II) to distinguish Laue classes within the same crystal system, where the lattice convention is the same as in previous versions of MAELAS\cite{maelas_publication2021,maelas_v2}. The second and third columns present the selected isotropic magnetoelastic constants ($b^{iso}$) and their corresponding convention, respectively. The fourth column gives the components of the parameterized strain tensor $\boldsymbol{\epsilon}(s)$, while the fifth shows the value of the parameter $\Gamma$  in Eq.\ref{eq:dE_tot}. The definition of each $b^{iso}$ is described in \ref{app_matrix}. The last two columns are related to the case including anisotropic magnetic interactions described by Eq.\ref{eq:dE_tot_ani}.}
\label{tab:method_data}
\resizebox{\textwidth}{!}{%
\begin{tabular}{@{}ccccccccc@{}}
\toprule
Crystal system  &
  \multicolumn{1}{c}{\begin{tabular}[c]{@{}c@{}}$b^{iso}$\end{tabular}}  &
  \multicolumn{1}{c}{Convention} &
  \multicolumn{1}{c}{$(\epsilon_{xx},\epsilon_{yy},\epsilon_{zz},\epsilon_{yz},\epsilon_{zx},\epsilon_{xy})$} &
  \multicolumn{1}{c}{$\Gamma$}  &
  \multicolumn{1}{c}{$\boldsymbol{\alpha}$} &
  \multicolumn{1}{c}{$\Lambda$} \\ \hline\hline
Cubic (I)      & \multicolumn{1}{c}{$b^{\alpha,2}$} &  \multicolumn{1}{c}{Lacheisserie\cite{booktremolet}}& \multicolumn{1}{c}{$(s,s,s,0,0,0)$} & \multicolumn{1}{c}{1} & $(0,0,1)$ & 0 \\ \hline
Hexagonal (I)   & \multicolumn{1}{c}{$b_{11}$}    & Clark\cite{CLARK1980531}                              &  $(s,s,0,0,0,0)$ & 2 & $(0,0,1)$ & $\frac{4}{3}b_{21}$  \\
      &    \multicolumn{1}{c}{$b_{12}$}    &  & \multicolumn{1}{c}{$(0,0,s,0,0,0)$}                              &  1  & $(0,0,1)$ & $\frac{2}{3}b_{22}$  \\ \hline
      Trigonal (I)   & \multicolumn{1}{c}{$b_{11}$}    & Cullen\cite{Cullen}                             &  $(s,s,0,0,0,0)$ & 2 & $(0,0,1)$ & $\frac{4}{3}b_{21}$  \\
      &    \multicolumn{1}{c}{$b_{12}$}    &  & \multicolumn{1}{c}{$(0,0,s,0,0,0)$}                              &  1 & $(0,0,1)$ & $\frac{2}{3}b_{22}$   \\ \hline
      Tetragonal (I)   & \multicolumn{1}{c}{$b_{11}$}    & Cullen\cite{Cullen} &  $(s,s,0,0,0,0)$ & 2 & $(0,0,1)$ & $\frac{4}{3}b_{21}$  \\
      &    \multicolumn{1}{c}{$b_{12}$}    &  & \multicolumn{1}{c}{$(0,0,s,0,0,0)$}                              &  1 & $(0,0,1)$ & $\frac{2}{3}b_{22}$  \\ \hline
Orthorhombic & \multicolumn{1}{c}{$b_{1}^{\alpha,0}$}    & Lacheisserie\cite{booktremolet}                             &  $(s,s,s,0,0,0)$ & 1 & $(0,0,1)$ & $\sqrt{2}b_{1}^{\alpha,2}$ \\
          & \multicolumn{1}{c}{$b_{2}^{\alpha,0}$}     & & $(-s/2,-s/2,s,0,0,0)$                             &  $\frac{1}{\sqrt{2}}$  & $(0,0,1)$ & $b_{2}^{\alpha,2}$ \\
            
          & \multicolumn{1}{c}{$b_{3}^{\alpha,0}$}     & & $(s,-s,0,0,0,0)$                             &  $\sqrt{\frac{2}{3}}$ & $(0,0,1)$ & $\sqrt{\frac{2}{3}}b_{3}^{\alpha,2}$ \\\bottomrule
\end{tabular}
}
\end{table}

\begin{table}[]
\centering
\caption{Calculated quantities with MAELAS v3.0 for each mode. Those quantities with -mode * mean that their definition does not fully decouple isotropic and anisotropic magnetic interactions, so that they cannot be calculated with a single mode implemented in MAELAS automatically. These quantities can be manually calculated by inserting the isotropic and anisotropic quantities obtained with MAELAS in the equations given in the Table \ref{tab:manual_data}. }
\label{tab:calculation_data}
\resizebox{\textwidth}{!}{%
\begin{tabular}{@{}c|cccc|cc|cc|cccc|cccc@{}}
\toprule
Crystal system  &
 \multicolumn{4}{c|}{Clark\cite{CLARK1980531} - Cullen\cite{Cullen}}   & \multicolumn{2}{c|}{Callen\cite{Callen}} & \multicolumn{2}{c|}{Birss\cite{Birss}} & \multicolumn{4}{c|}{Mason\cite{Mason}-Nieves\cite{maelas_publication2021}} & \multicolumn{4}{c}{Lacheisserie\cite{booktremolet}} \\ \hline
  Space Group & \multicolumn{1}{c}{$b$} & \multicolumn{1}{c}{-mode} & \multicolumn{1}{c}{$\lambda$} &  \multicolumn{1}{c|}{-mode}  & \multicolumn{1}{c}{$\lambda$} &  \multicolumn{1}{c|}{-mode}  & \multicolumn{1}{c}{$\lambda$} &  \multicolumn{1}{c|}{-mode} & \multicolumn{1}{c}{$b$} & \multicolumn{1}{c}{-mode} & \multicolumn{1}{c}{$\lambda$} &  \multicolumn{1}{c|}{-mode} & \multicolumn{1}{c}{$b$} & \multicolumn{1}{c}{-mode} & \multicolumn{1}{c}{$\lambda$} &  \multicolumn{1}{c}{-mode} \\\hline
Cubic (I)      & \multicolumn{1}{c}{$b_{0}$} &  \multicolumn{1}{c}{*}& \multicolumn{1}{c}{$\lambda^{\alpha}$} & \multicolumn{1}{c|}{3} &  &  & & &  &  & & & \multicolumn{1}{c}{$b^{\alpha,2}$} & 3 & \multicolumn{1}{c}{$\lambda^{\alpha,2}$} & 3 \\
  SG 207-230       &   \multicolumn{1}{c}{$b_{1}$} & \multicolumn{1}{c}{1,2} & \multicolumn{1}{c}{$\lambda_{001}$} & \multicolumn{1}{c|}{1,2} &  &  & & &  &  & & &  \multicolumn{1}{c}{$b^{\gamma,2}$} & 1,2 & \multicolumn{1}{c}{$\lambda^{\gamma,2}$} & 1,2 \\
         &   \multicolumn{1}{c}{$b_{2}$} & \multicolumn{1}{c}{1,2} & \multicolumn{1}{c}{$\lambda_{111}$} & \multicolumn{1}{c|}{1,2} &  &  & & &  &  & & &  \multicolumn{1}{c}{$b^{\epsilon,2}$} & 1,2 & \multicolumn{1}{c}{$\lambda^{\epsilon,2}$} & 1,2 \\ \hline
Hexagonal (I)   & \multicolumn{1}{c}{$b_{11}$}    & \multicolumn{1}{c}{3}                              &  \multicolumn{1}{c}{$\lambda^{\alpha1,0}$} & \multicolumn{1}{c|}{3} &   \multicolumn{1}{c}{$\lambda^{\alpha}_{11}$} &  *  &  $Q_0$ & * &  & & \multicolumn{1}{c}{$\lambda^{\alpha1,0}$} & * &  \multicolumn{1}{c}{$b_1^{\alpha,0}$} & 3 & \multicolumn{1}{c}{$\lambda_1^{\alpha,0}$} & 3  \\
       SG 177-194  &    \multicolumn{1}{c}{$b_{12}$}    & \multicolumn{1}{c}{3}                              &  \multicolumn{1}{c}{$\lambda^{\alpha2,0}$} & \multicolumn{1}{c|}{3} &  \multicolumn{1}{c}{$\lambda^{\alpha}_{21}$} & 3 & $Q_1$ & * &  & & \multicolumn{1}{c}{$\lambda^{\alpha2,0}$} & * &  \multicolumn{1}{c}{$b_2^{\alpha,0}$} & 3 & \multicolumn{1}{c}{$\lambda_2^{\alpha,0}$} & 3 \\
          & \multicolumn{1}{c}{$b_{21}$}    & \multicolumn{1}{c}{1,2}                              &  \multicolumn{1}{c}{$\lambda^{\alpha1,2}$} & \multicolumn{1}{c|}{1,2} &  \multicolumn{1}{c}{$\lambda^{\alpha}_{12}$} &  1,2 &  $Q_2$ & 1,2 &  & & \multicolumn{1}{c}{$\lambda_A$} & 1,2 &  \multicolumn{1}{c}{$b_1^{\alpha,2}$} & 1,2 & \multicolumn{1}{c}{$\lambda_1^{\alpha,2}$} & 1,2 \\
         &    \multicolumn{1}{c}{$b_{22}$}    & \multicolumn{1}{c}{1,2}                              &  \multicolumn{1}{c}{$\lambda^{\alpha2,2}$} & \multicolumn{1}{c|}{1,2} &  \multicolumn{1}{c}{$\lambda^{\alpha}_{22}$} & 1,2 &   $Q_4$ & 1,2 &  & & \multicolumn{1}{c}{$\lambda_B$} & 1,2 &  \multicolumn{1}{c}{$b_2^{\alpha,2}$} & 1,2 & \multicolumn{1}{c}{$\lambda_2^{\alpha,2}$} & 1,2 \\
          & \multicolumn{1}{c}{$b_{3}$}    & \multicolumn{1}{c}{1,2}                              &  \multicolumn{1}{c}{$\lambda^{\gamma,2}$} & \multicolumn{1}{c|}{1,2} &  \multicolumn{1}{c}{$\lambda^{\gamma}$} & 1,2  & $Q_6$ & 1,2 &  & & \multicolumn{1}{c}{$\lambda_C$} & 1,2 &  \multicolumn{1}{c}{$b^{\epsilon,2}$} & 1,2 & \multicolumn{1}{c}{$\lambda^{\epsilon,2}$} & 1,2 \\
        & \multicolumn{1}{c}{$b_{4}$}    & \multicolumn{1}{c}{1,2}                              &  \multicolumn{1}{c}{$\lambda^{\epsilon,2}$} & \multicolumn{1}{c|}{1,2} &  \multicolumn{1}{c}{$\lambda^{\epsilon}$} & 1,2  & $Q_8$ & 1,2 &  & & \multicolumn{1}{c}{$\lambda_D$} & 1,2 &  \multicolumn{1}{c}{$b^{\zeta,2}$} & 1,2 & \multicolumn{1}{c}{$\lambda^{\zeta,2}$} & 1,2 \\ \hline
Trigonal (I)   & \multicolumn{1}{c}{$b_{11}$}    & \multicolumn{1}{c}{3}                              &  \multicolumn{1}{c}{$\lambda^{\alpha1,0}$} & \multicolumn{1}{c|}{3} &  &  &  &  &  &  & & &  & & & \\
   SG 149-167       &    \multicolumn{1}{c}{$b_{12}$}   & \multicolumn{1}{c}{3}   & \multicolumn{1}{c}{$\lambda^{\alpha2,0}$}                              &  \multicolumn{1}{c|}{3} &  &  &  & &  &  & & &  & & &  \\
    & \multicolumn{1}{c}{$b_{21}$}    & \multicolumn{1}{c}{1,2}                              &  \multicolumn{1}{c}{$\lambda^{\alpha1,2}$} & \multicolumn{1}{c|}{1,2} &  & &  & &  &  & & &  & & & \\
       &    \multicolumn{1}{c}{$b_{22}$}   & \multicolumn{1}{c}{1,2}   & \multicolumn{1}{c}{$\lambda^{\alpha2,2}$}                              &  \multicolumn{1}{c|}{1,2} &  & &  & &  &  & & &  & & & \\
         & \multicolumn{1}{c}{$b_{3}$}    & \multicolumn{1}{c}{1,2}                              &  \multicolumn{1}{c}{$\lambda^{\gamma1}$} & \multicolumn{1}{c|}{1,2} &  & &  & &  &  & & &  & & &  \\
         & \multicolumn{1}{c}{$b_{4}$}    & \multicolumn{1}{c}{1,2}                              &  \multicolumn{1}{c}{$\lambda^{\gamma2}$} & \multicolumn{1}{c|}{1,2} &  & &  &  &  &  & & &  & & &  \\
           & \multicolumn{1}{c}{$b_{14}$}    & \multicolumn{1}{c}{1,2}                              &  \multicolumn{1}{c}{$\lambda_{12}$} & \multicolumn{1}{c|}{1,2} &  & &  & &  &  & & &  & & &  \\
            & \multicolumn{1}{c}{$b_{34}$}    & \multicolumn{1}{c}{1,2}                              &  \multicolumn{1}{c}{$\lambda_{21}$} & \multicolumn{1}{c|}{1,2} &  & &  &  &  &  & & &  & & & \\ \hline
Tetragonal (I) &  \multicolumn{1}{c}{$b_{11}$}    & \multicolumn{1}{c}{3}                              &  \multicolumn{1}{c}{$\lambda^{\alpha1,0}$} & \multicolumn{1}{c|}{3} & &  & & &  &  & \multicolumn{1}{c}{$\lambda^{\alpha1,0}$} &  * &  \multicolumn{1}{c}{$b_1^{\alpha,0}$} & 3 & \multicolumn{1}{c}{$\lambda_1^{\alpha,0}$} & 3 \\
    SG 89-142       & \multicolumn{1}{c}{$b_{12}$}                     & \multicolumn{1}{c}{3}          &  \multicolumn{1}{c}{$\lambda^{\alpha2,0}$} & \multicolumn{1}{c|}{3} &  & &  & &  &  & \multicolumn{1}{c}{$\lambda^{\alpha2,0}$} & * &  \multicolumn{1}{c}{$b_2^{\alpha,0}$} & 3 & \multicolumn{1}{c}{$\lambda_2^{\alpha,0}$} & 3 \\
    &  \multicolumn{1}{c}{$b_{21}$}    & \multicolumn{1}{c}{1,2}                              &  \multicolumn{1}{c}{$\lambda^{\alpha1,2}$} & \multicolumn{1}{c|}{1,2} &  & &  &  &  &  & \multicolumn{1}{c}{$\lambda_1$} & 1,2 &  \multicolumn{1}{c}{$b_1^{\alpha,2}$} & 1,2 & \multicolumn{1}{c}{$\lambda_1^{\alpha,2}$} & 1,2 \\
          & \multicolumn{1}{c}{$b_{22}$}                     & \multicolumn{1}{c}{1,2}          &  \multicolumn{1}{c}{$\lambda^{\alpha2,2}$} & \multicolumn{1}{c|}{1,2} &  & &  & &  &  & \multicolumn{1}{c}{$\lambda_2$} & 1,2 &  \multicolumn{1}{c}{$b_2^{\alpha,2}$} & 1,2 & \multicolumn{1}{c}{$\lambda_2^{\alpha,2}$} & 1,2 \\
          & \multicolumn{1}{c}{$b_{3}$}    & \multicolumn{1}{c}{1,2}                              &  \multicolumn{1}{c}{$\lambda^{\gamma,2}$} & \multicolumn{1}{c|}{1,2} &  & &  & &  &  & \multicolumn{1}{c}{$\lambda_3$} & 1,2 &  \multicolumn{1}{c}{$b^{\gamma,2}$} & 1,2 & \multicolumn{1}{c}{$\lambda^{\gamma,2}$} & 1,2 \\
           & \multicolumn{1}{c}{$b_{4}$}    & \multicolumn{1}{c}{1,2}                              &  \multicolumn{1}{c}{$\lambda^{\epsilon,2}$} & \multicolumn{1}{c|}{1,2} & &  & & &  &  & \multicolumn{1}{c}{$\lambda_4$} & 1,2 & \multicolumn{1}{c}{$b^{\delta,2}$} & 1,2 & \multicolumn{1}{c}{$\lambda^{\delta,2}$} & 1,2 \\
         & \multicolumn{1}{c}{$b'_{3}$}    & \multicolumn{1}{c}{1,2}                              &  \multicolumn{1}{c}{$\lambda^{\delta,2}$} & \multicolumn{1}{c|}{1,2} &  & &  & &  &  &\multicolumn{1}{c}{$\lambda_5$} & 1,2 &  \multicolumn{1}{c}{$b^{\epsilon,2}$} & 1,2 & \multicolumn{1}{c}{$\lambda^{\epsilon,2}$} & 1,2 \\ \hline
Orthorhombic &  &                               &   &  &  & &  & & $b_{01} $ & * & \multicolumn{1}{c}{$\lambda^{\alpha1,0}$}  & * &  \multicolumn{1}{c}{$b_1^{\alpha,0}$} & 3 & \multicolumn{1}{c}{$\lambda_1^{\alpha,0}$} & 3 \\
  SG 16-74      &    &                            &  &  &  & &  & &  $b_{02} $ & * & \multicolumn{1}{c}{$\lambda^{\alpha2,0}$}  & * &  \multicolumn{1}{c}{$b_2^{\alpha,0}$} & 3 & \multicolumn{1}{c}{$\lambda_2^{\alpha,0}$} & 3 \\
            &     &                               &   &  & &  & & &  $b_{03} $ & * & \multicolumn{1}{c}{$\lambda^{\alpha3,0}$}  & * &  \multicolumn{1}{c}{$b_3^{\alpha,0}$} & 3 & \multicolumn{1}{c}{$\lambda_3^{\alpha,0}$} & 3 \\
            &     &                               &   &  & &  & & &  $b_{1} $ & 1,2 & \multicolumn{1}{c}{$\lambda_1$}  & 1,2 &  \multicolumn{1}{c}{$b_1^{\alpha,2}$} & 1,2 & \multicolumn{1}{c}{$\lambda_1^{\alpha,2}$} & 1,2 \\
            &     &                               &   &  & &  & & &   $b_{2} $ & 1,2 & \multicolumn{1}{c}{$\lambda_2$}  & 1,2 &  \multicolumn{1}{c}{$b_2^{\alpha,2}$} & 1,2 & \multicolumn{1}{c}{$\lambda_2^{\alpha,2}$} & 1,2 \\
            &     &                               &   &  & &  & & &   $b_{3} $ & 1,2 & \multicolumn{1}{c}{$\lambda_3$}  & 1,2 &  \multicolumn{1}{c}{$b_3^{\alpha,2}$} & 1,2 & \multicolumn{1}{c}{$\lambda_3^{\alpha,2}$} & 1,2 \\
            &     &                               &   &  & &  & & &   $b_{4} $ & 1,2 & \multicolumn{1}{c}{$\lambda_4$}  & 1,2 &  \multicolumn{1}{c}{$b_1^{\alpha,2'}$} & 1,2 & \multicolumn{1}{c}{$\lambda_1^{\alpha,2'}$} & 1,2 \\
            &     &                               &   &  & &  & & &   $b_{5} $ & 1,2 & \multicolumn{1}{c}{$\lambda_5$}  & 1,2 &  \multicolumn{1}{c}{$b_2^{\alpha,2'}$} & 1,2 & \multicolumn{1}{c}{$\lambda_2^{\alpha,2'}$} & 1,2 \\
            &     &                               &   &  & &  & & &   $b_{6} $ & 1,2 & \multicolumn{1}{c}{$\lambda_6$}  & 1,2 & \multicolumn{1}{c}{$b_3^{\alpha,2'}$} & 1,2 & \multicolumn{1}{c}{$\lambda_3^{\alpha,2'}$} & 1,2 \\
            &     &                               &   &  & &  & & &   $b_{7} $ & 1,2 & \multicolumn{1}{c}{$\lambda_7$}  & 1,2 &  \multicolumn{1}{c}{$b^{\beta,2}$} & 1,2 & \multicolumn{1}{c}{$\lambda^{\beta,2}$} & 1,2 \\
            &     &                               &   &  & &  & & &   $b_{8} $ & 1,2 & \multicolumn{1}{c}{$\lambda_8$}  & 1,2 &  \multicolumn{1}{c}{$b^{\gamma,2}$} & 1,2 & \multicolumn{1}{c}{$\lambda^{\gamma,2}$} & 1,2 \\
            &     &                               &   &  & &  & & &   $b_{9} $ & 1,2 & \multicolumn{1}{c}{$\lambda_9$}  & 1,2 &  \multicolumn{1}{c}{$b^{\delta,2}$} & 1,2 & \multicolumn{1}{c}{$\lambda^{\delta,2}$} & 1,2 \\
             \bottomrule
\end{tabular}
}
\end{table}

\begin{table}[]
\centering
\caption{The definition of the quantities shown in the third column does not fully decouple isotropic and anisotropic interactions, so that they cannot be calculated with a single mode implemented in MAELAS automatically. These quantities can be manually calculated by inserting the isotropic and anisotropic quantities obtained with MAELAS in the equations given in the fourth column.}
\label{tab:manual_data}
\resizebox{\textwidth}{!}{%
\begin{tabular}{@{}c|cc|cc@{}}
\toprule
Crystal system  &
  \multicolumn{1}{c}{Convention} &
  * &
  = &  \multicolumn{1}{c}{Convention} \\ \hline\hline
Cubic (I)      & \multicolumn{1}{c}{Clark-Cullen} &  \multicolumn{1}{c|}{$b_0$} & \multicolumn{1}{c}{$\frac{1}{3}(b^{\alpha,2}-b^{\gamma,2})$} & \multicolumn{1}{c}{Lacheisserie} \\ \hline
Hexagonal (I)      & \multicolumn{1}{c}{Callen} &  \multicolumn{1}{c|}{$\lambda_{11}^\alpha$} & \multicolumn{1}{c}{$2\lambda^{\alpha1,0}+\lambda^{\alpha2,0}+2\lambda^{\alpha1,2}+\lambda^{\alpha2,2}$} & \multicolumn{1}{c}{Clark-Cullen} \\
    & \multicolumn{1}{c}{Birss} &  \multicolumn{1}{c|}{$Q_0$} & \multicolumn{1}{c}{$\lambda^{\alpha1,0}+\frac{2}{3}\lambda^{\alpha1,2}$} & \multicolumn{1}{c}{Clark-Cullen} \\
     & \multicolumn{1}{c}{Birss} &  \multicolumn{1}{c|}{$Q_1$} & \multicolumn{1}{c}{$\lambda^{\alpha2,0}+\frac{2}{3}\lambda^{\alpha2,2}-\lambda^{\alpha1,0}-\frac{2}{3}\lambda^{\alpha1,2}$} & \multicolumn{1}{c}{Clark-Cullen} \\
     & \multicolumn{1}{c}{Mason-Nieves} &  \multicolumn{1}{c|}{$\lambda^{\alpha1,0}$} & \multicolumn{1}{c}{$\lambda^{\alpha1,0}+\frac{2}{3}\lambda^{\alpha1,2}$} & \multicolumn{1}{c}{Clark-Cullen} \\
     & \multicolumn{1}{c}{Mason-Nieves} &  \multicolumn{1}{c|}{$\lambda^{\alpha2,0}$} & \multicolumn{1}{c}{$\lambda^{\alpha2,0}+\frac{2}{3}\lambda^{\alpha2,2}$} & \multicolumn{1}{c}{Clark-Cullen} \\\hline
     Tetragonal (I) & \multicolumn{1}{c}{Mason-Nieves} &  \multicolumn{1}{c|}{$\lambda^{\alpha1,0}$} & \multicolumn{1}{c}{$\lambda^{\alpha1,0}+\frac{2}{3}\lambda^{\alpha1,2}$} & \multicolumn{1}{c}{Clark-Cullen} \\
     & \multicolumn{1}{c}{Mason-Nieves} &  \multicolumn{1}{c|}{$\lambda^{\alpha2,0}$} & \multicolumn{1}{c}{$\lambda^{\alpha2,0}+\frac{2}{3}\lambda^{\alpha2,2}$} & \multicolumn{1}{c}{Clark-Cullen} \\
    \hline
    Orthorhombic & \multicolumn{1}{c}{Mason-Nieves} &  \multicolumn{1}{c|}{$b_{01}$} & \multicolumn{1}{c}{$\frac{1}{3}(b_1^{\alpha,0} + \sqrt{2} b_1^{\alpha,2}-\frac{1}{\sqrt{2}} b_2^{\alpha,0} - b_2^{\alpha,2}) + \frac{1}{\sqrt{6}} (b_3^{\alpha,0} + b_3^{\alpha,2})$} & \multicolumn{1}{c}{Lacheisserie} \\
     & \multicolumn{1}{c}{Mason-Nieves} &  \multicolumn{1}{c|}{$b_{02}$} & \multicolumn{1}{c}{$\frac{1}{3}(b_1^{\alpha,0} + \sqrt{2} b_1^{\alpha,2}-\frac{1}{\sqrt{2}} b_2^{\alpha,0} - b_2^{\alpha,2}) - \frac{1}{\sqrt{6}} (b_3^{\alpha,0} + b_3^{\alpha,2})$} & \multicolumn{1}{c}{Lacheisserie} \\
     & \multicolumn{1}{c}{Mason-Nieves} &  \multicolumn{1}{c|}{$b_{03}$} & \multicolumn{1}{c}{$\frac{1}{3}(b_1^{\alpha,0} + \sqrt{2} b_1^{\alpha,2}+\sqrt{2} b_2^{\alpha,0} + 2b_2^{\alpha,2})$} & \multicolumn{1}{c}{Lacheisserie} \\
     & \multicolumn{1}{c}{Mason-Nieves} &  \multicolumn{1}{c|}{$\lambda^{\alpha1,0}$} & \multicolumn{1}{c}{$\frac{1}{3} (\lambda_1^{\alpha,0} + \lambda_1^{\alpha,2} - \lambda_2^{\alpha,0} - \lambda_2^{\alpha,2}) + \lambda_3^{\alpha,0} + \lambda_3^{\alpha,2}$} & \multicolumn{1}{c}{Lacheisserie} \\
     & \multicolumn{1}{c}{Mason-Nieves} &  \multicolumn{1}{c|}{$\lambda^{\alpha2,0}$} & \multicolumn{1}{c}{$\frac{1}{3} (\lambda_1^{\alpha,0} + \lambda_1^{\alpha,2} - \lambda_2^{\alpha,0} - \lambda_2^{\alpha,2}) - \lambda_3^{\alpha,0} - \lambda_3^{\alpha,2}$} & \multicolumn{1}{c}{Lacheisserie} \\
     & \multicolumn{1}{c}{Mason-Nieves} &  \multicolumn{1}{c|}{$\lambda^{\alpha3,0}$} & \multicolumn{1}{c}{$\frac{1}{3} (\lambda_1^{\alpha,0} + \lambda_1^{\alpha,2} + 2\lambda_2^{\alpha,0} +2 \lambda_2^{\alpha,2})$} & \multicolumn{1}{c}{Lacheisserie}
\\\bottomrule
\end{tabular}
}
\end{table}

\subsection{Including anisotropic magnetic interactions}
\label{subsection:method_ani}

The methodology described in Section \ref{subsection:method_iso} should be slightly modified in order to take into account the anisotropic effects. Firstly, if we include anisotropic magnetic interactions, then the total energy now reads
\begin{equation}
    E(\boldsymbol{\epsilon},\boldsymbol{\alpha})=E_{el}(\boldsymbol{\epsilon})+E_{me}(\boldsymbol{\epsilon},\boldsymbol{\alpha})+E_{K}^{0}(\boldsymbol{\alpha}),
\label{eq:E_tot_ani}
\end{equation}
where  $\boldsymbol{\alpha}$ is the normalized magnetization vector ($\vert\boldsymbol{\alpha}\vert =1$), $E_{K}^{0}$ is the unstrained magnetocrystalline anisotropy energy, and $E_{me}(\boldsymbol{\epsilon},\boldsymbol{\alpha})=E_{me}^{iso}(\boldsymbol{\epsilon})+E_{me}^{ani}(\boldsymbol{\epsilon},\boldsymbol{\alpha})$ contains only linear terms in the strain up to second order in the magnetization direction $\boldsymbol{\alpha}$. Next, if we constrain the magnetization direction along z-axis ($\boldsymbol{\alpha}=(0,0,1)$), and we apply the parameterized strain tensor $\boldsymbol{\epsilon}(s)$ given in Table \ref{tab:method_data} for the i-th isotropic magnetoelastic constant $b_{i}^{iso}$, then Eq.\ref{eq:E_tot_ani} becomes  
\begin{equation}
   \frac{1}{V_0} E(\boldsymbol{\epsilon}^i(s),\boldsymbol{\alpha}=(0,0,1))=\Phi_i(C_{ij})s^2+[\Gamma_i b^{iso}_i+\Lambda_i(b^{ani})] s + \frac{1}{V_0}E_0,
\label{eq:dE_tot_ani}
\end{equation}
where $\Lambda_i(b^{ani})$ depends on the anisotropic magnetoelastic constants and its value is shown Table \ref{tab:method_data}. The value of $\Gamma_i$ is the same as in Eq.\ref{eq:dE_tot}. Fitting Eq.\ref{eq:dE_tot_ani} to a third order polynomial (Eq.\ref{eq:dE_tot_fit}) allows
to derive the i-th isotropic magnetoelastic constant $b_i^{iso}$ via
\begin{equation}
    b_i^{iso}=\frac{\mathcal{C}-\Lambda_i(b^{ani})}{\Gamma_i},
\label{eq:b_i_sol_ani}
\end{equation}
where $\mathcal{C}$ is a fitting parameter. Here,  the anisotropic magnetoelastic constants in $\Lambda_i(b^{ani})$ could be obtained through -mode 2 (which is more accurate than -mode 1\cite{maelas_v2}) using the same reference state as in -mode 3. Note for the cubic (I) symmetry the anisotropic magnetic interactions gives a null correction $\Lambda(b^{ani})=0$, see Table \ref{tab:method_data}. To include anisotropic magnetic interactions in -mode 3, the user needs to add the flag -ani in Steps 3 and 5 (see Fig.\ref{fig:workflow_1}). Moreover, in Step 5 one also needs to have the file called MAGANI generated in Step 5 of -mode 2  which contains the data of the anisotropic magnetoelastic constants $b^{ani}$, required to compute $\Lambda_i(b^{ani})$.

\section{Computational details}
\label{section:comp_details}

\begin{figure}[h!]
\centering
\includegraphics[width=\columnwidth ,angle=0]{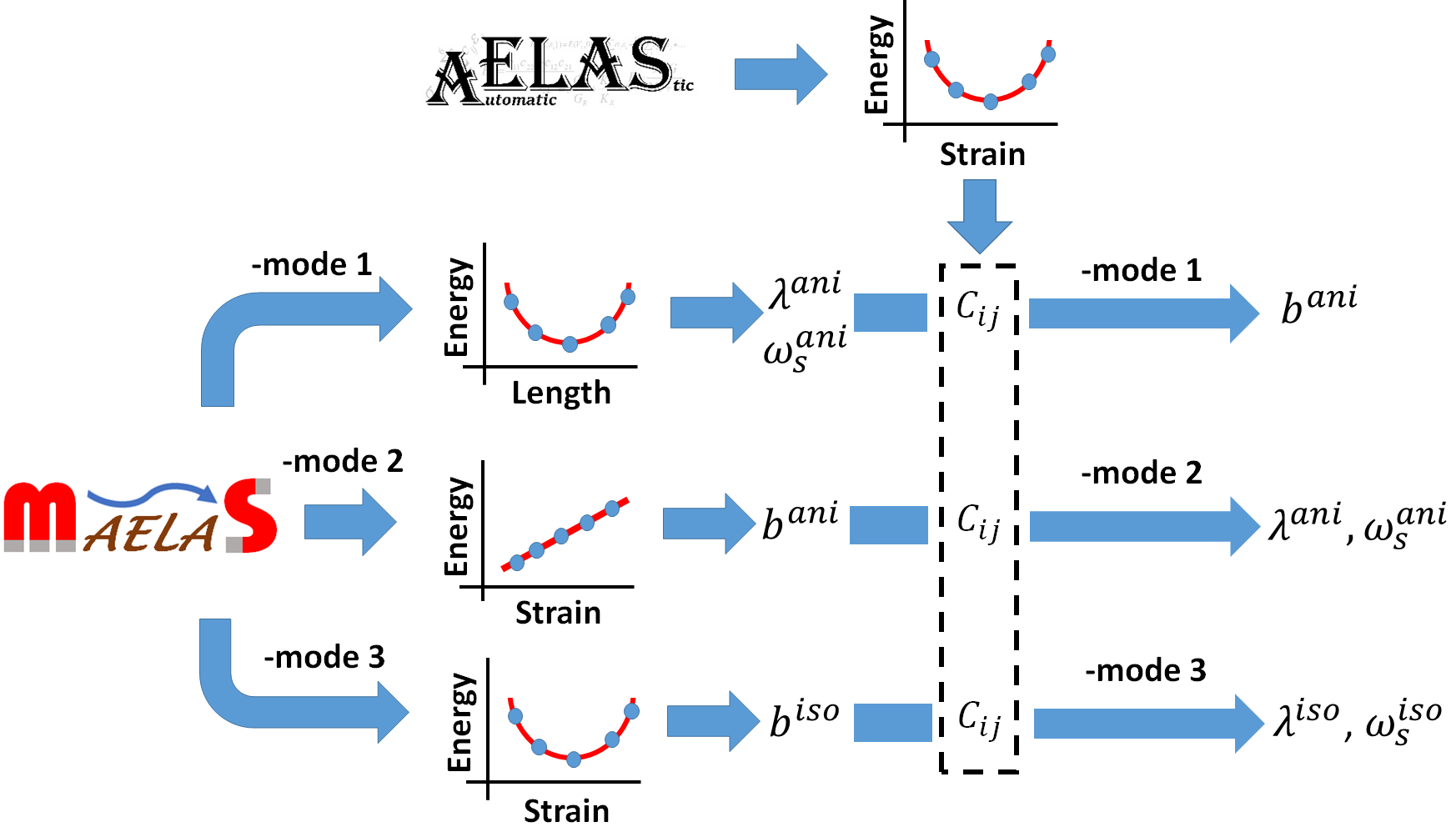}
\caption{Diagram showing the three methods available in MAELAS v3.0.}
\label{fig:diagram_3_modes}
\end{figure}

\begin{figure}[h!]
\centering
\includegraphics[width=\columnwidth ,angle=0]{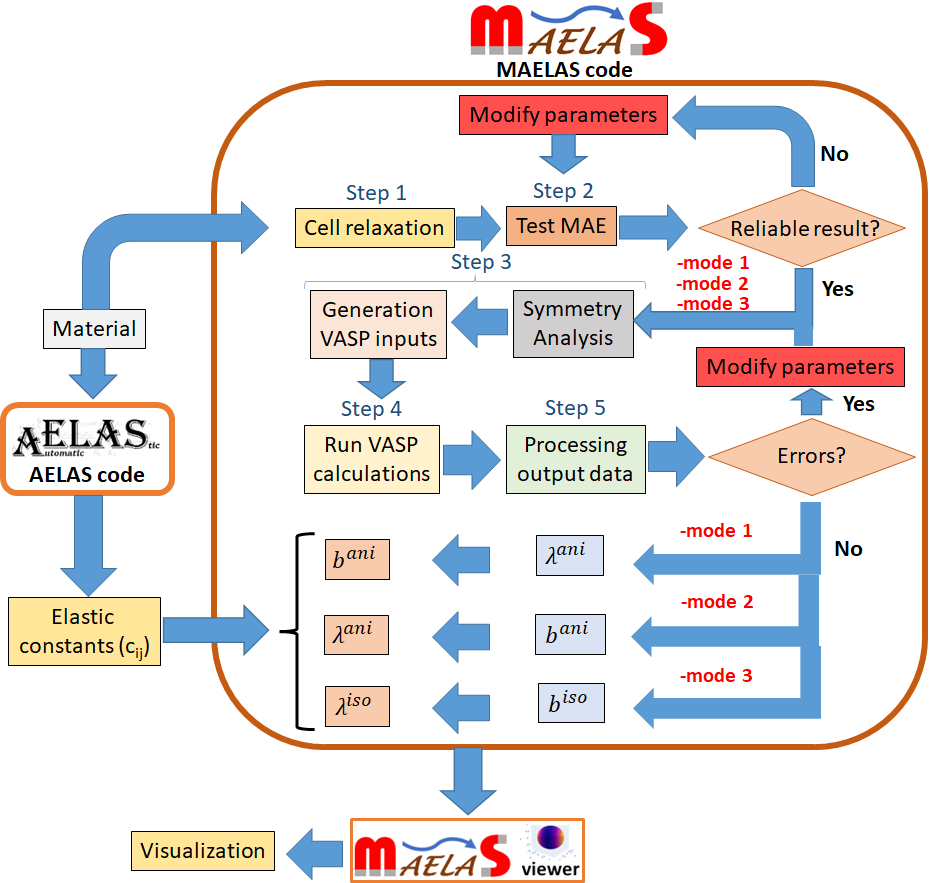}
\caption{General workflow of the methods implemented in MAELAS v3.0.}
\label{fig:workflow_1}
\end{figure}

We have implemented the methodology described in Section \ref{section:method} in the new version of MAELAS v3.0, which can be executed by using the flag -mode 3 in command line. This new method complements the previous available modes (-mode 1 and -mode 2) associated with the anisotropic quantities, see Fig.\ref{fig:diagram_3_modes}. The corresponding workflow for this new methodology is depicted in Fig.\ref{fig:workflow_1}. It contains the same steps as the workflow for the methods implemented in previous versions for the calculation of anisotropic magnetoelastic constants and magnetostrictive coefficients (-mode 1 and -mode 2) \cite{maelas_publication2021,maelas_v2}. However, for -mode 3 the output from the cell relaxation (step 1) should correspond to the reference state described in Section \ref{subsection:method_iso}. Note that the test of the magnetocrystalline anisotropy (step 2) is not necessary if SOC is not included, but it still could be useful to verify the isotropic magnetic character without SOC. The generated deformations for the Vienna Ab initio Simulation Package (VASP)  \cite{vasp_1,vasp_2,vasp_3} in step 3 corresponds to the parameterized strain tensor shown in Table \ref{tab:method_data}. By default, -mode 3 assumes a FM GS generating FM configurations for the input files required for the calculation of the energy of the deformed unit cells. If the GS is AFM, then the generated FM configurations should be changed to AFM manually. In step 5, after processing VASP output data, -mode 3 of MAELAS v3.0 will calculate directly the isotropic magnetoelastic constants. Additionally, if the elastic constants are also provided in AELAS format \cite{AELAS}, then it will also calculate the isotropic magnetostrictive coefficients and isotropic contribution to spontaneous volume magnetostriction from the theoretical equations ($\lambda^{iso}(b^{iso}_{k},C_{ij})$ and $\omega^{iso}_s(\lambda^{iso})$) given in \ref{app_matrix}. The anisotropic contribution to spontaneous volume magnetostriction ($\omega^{ani}_s$) can be computed via -mode 1 or -mode 2, see \ref{app_matrix}. The calculated magnetostrictive coefficients can be analyzed and visualized with the online tool MAELASviewer \cite{maelasviewer}. In Table \ref{tab:calculation_data}, we show all calculated quantities with MAELAS v3.0 for each mode. Those quantities with -mode * in Table \ref{tab:calculation_data} mean that their definition does not fully decouple isotropic and anisotropic interactions, so that they cannot be calculated with a single mode implemented in MAELAS automatically. These quantities can be manually calculated by inserting the isotropic and anisotropic quantities obtained with MAELAS in the equations given in the Table \ref{tab:manual_data}. The -mode 3 is quite robust for all supported crystal symmetries since is based on the energy-strain method, exhibiting very small relative errors like in -mode 2\cite{maelas_v2}. The source files of the analysis of the accuracy of -mode 3 can be found in Example folder\cite{maelasexample}.

By default, in step 3 MAELAS applies a conventional cell transformation to the input unit cell in this step. After this transformation, it generates the deformed unit cells. This is done to ensure the same lattice IEEE convention as in AELAS. Note that this transformation might change the size, number of atoms and reference axis of the original unit cell. Reference axis in the calculated quantities by MAELAS  corresponds to the one used in the transformed/deformed unit cells (output from step 3) not to the input one in step 3. The conventional cell transformation can be avoided by adding flag -nc both in steps 3 and 5. This option could be helpful for supercells and SQS, but in this case one should carefully verify the consistency with the calculation of elastic constants with AELAS. In this new release of MAELAS, we solved some issues found in previous versions, so that we recommend to use the latest version available in GitHub repository\cite{Maelas}.

\section{Tests for the new method}
\label{section:test}

In this section, we apply the presented methodology in Section \ref{section:method} (-mode 3) to classical spin-lattice models and first-principles calculations. In the case of classical spin-lattice models, this is achieved by an interface between MAELAS and the SPIN package of LAMMPS\cite{TRANCHIDA2018406,thompson2022lammps}, while MAELAS is interfaced with VASP\cite{vasp_1,vasp_2,vasp_3} in the case of first-principles calculations. Before using first-principle calculations, it is convenient to apply this method to a simple classical spin-lattice model \cite{TRANCHIDA2018406,nieves2021spinlattice_prb}. The simplicity of this kind of models can help to verify the correct implementation of the method in MAELAS and its level of accuracy easily, as well as to provide a deep understanding of the physics behind the proposed methodology. For example, within these models one can compute the equilibrium volume at the zero-temperature PM state ($V_{PM}$) quite accurately and rapidly\cite{nieves2021spinlattice_prb}, as well as derive an exact theoretical expression for $\omega_s$\cite{Chika,nieves2021spinlattice_prb}, offering  additional ways to verify MAELAS methodology.


\subsection{Classical spin-lattice model}
\label{subsection:lmp}

 For the application of MAELAS to a classical spin-lattice model, we consider the Hamiltonian 
\begin{equation}
\begin{aligned}
\mathcal{H}_{sl}(\boldsymbol{r},\boldsymbol{p},\boldsymbol{s}) & =  \mathcal{H}_{mag}(\boldsymbol{r},\boldsymbol{s})+\sum_{i=1}^N\frac{\boldsymbol{p}_i}{2m_i}+\sum_{i,j=1}^N\mathcal{V}(r_{ij}),
\label{eq:Ham_tot}
\end{aligned}
\end{equation}
where $\boldsymbol{r}_i$, $\boldsymbol{p}_i$, $\boldsymbol{s}_i$, and $m_i$ stand for the position, momentum, normalized magnetic moment and mass for each atom $i$ in the system, respectively, $\mathcal{V}(r_{ij})=\mathcal{V}(\vert \boldsymbol{r}_i-\boldsymbol{r}_j\vert)$ is the interatomic potential energy and $N$ is the total number of atoms in the system with total volume $V$. For the sake of simplicity, we only include the exchange interactions in the magnetic energy ($\mathcal{H}_{mag}$), that is
\begin{equation}
\begin{aligned}
\mathcal{H}_{mag}(\boldsymbol{r},\boldsymbol{s}) & = -\frac{1}{2}\sum_{i,j=1,i\neq j}^N \Tilde{J}(r_{ij})\boldsymbol{\mu}_i(r_{ij})\cdot\boldsymbol{\mu}_j(r_{ij}) = -\frac{1}{2}\sum_{i,j=1,i\neq j}^N J(r_{ij})\boldsymbol{s}_i\cdot\boldsymbol{s}_j,
\label{eq:Ham_mag}
\end{aligned}
\end{equation}
where $\boldsymbol{\mu}_i$ is the atomic magnetic moment vector for each atom $i$ in the system, $J(r_{ij})=\Tilde{J}(r_{ij})\mu_i(r_{ij})\mu_j(r_{ij})$ is the exchange parameter. We parameterize $J(r_{ij})$ using a single Bethe-Slater function, as implemented in the SPIN package of LAMMPS \cite{TRANCHIDA2018406}
\begin{equation}
\begin{aligned}
J(r_{ij}) & =  4\alpha \left(\frac{r_{ij}}{\delta}\right)^2 \left[1-\gamma\left(\frac{r_{ij}}{\delta}\right)^2 \right]e^{-\left(\frac{r_{ij}}{\delta}\right)^2} \Theta(R_{c}-r_{ij}),
\label{eq:BS_J}
\end{aligned}
\end{equation}
where $\Theta(R_{c}-r_{ij})$ is the Heaviside step function and $R_{c}$ is the cut-off radius and $\alpha$, $\gamma$, and $\delta$ are parameters. The value of these parameters are set to $\alpha=27.26$ meV,  $\gamma=0.2171$ and $\delta=1.841$ \r{A} similar to Ref.\cite{TRANCHIDA2018406} for bcc Fe. Here, we restrict the exchange interaction up to second nearest neighbors by setting the cut-off parameter to $R_c=3.5$ \r{A}. For the classical interatomic potential $\mathcal{V}(r_{ij})$ we use the embedded-atom method (EAM) potential for bcc Fe developed by Chamati et al.\cite{CHAMATI20061793}. Below, we study this model with three different approaches: i) atomistic simulations, ii) MAELAS, and iii) theory. In Section \ref{subsection:model_lmp}, we discuss about possible extensions of this spin-lattice model.

\begin{figure}[h!]
\centering
\includegraphics[width=\columnwidth ,angle=0]{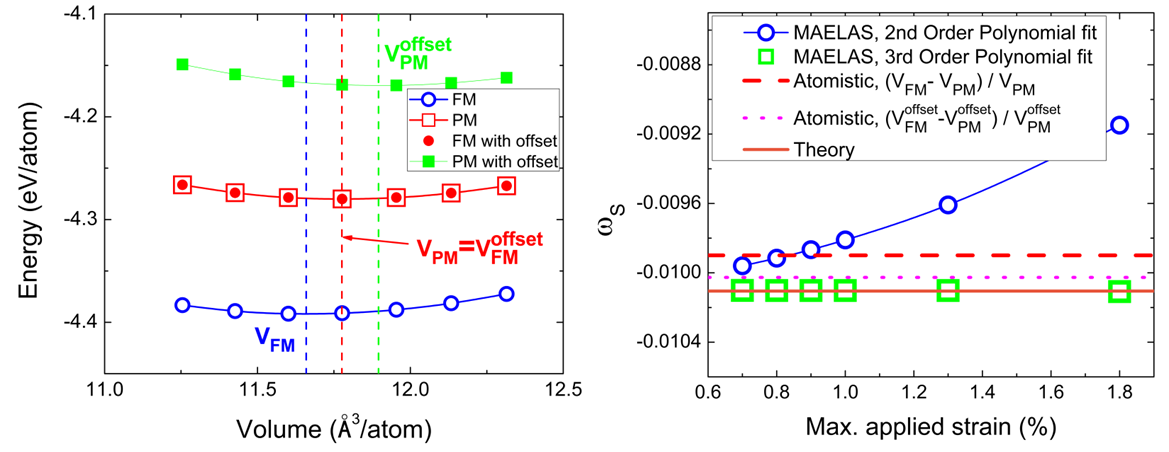}
\caption{(Left) Calculated EOS using atomistic simulations  for the classical spin-lattice model of bcc Fe described in Section \ref{subsection:lmp}. Open blue circles and open red squares represent the FM and PM states without offset in the exchange energy, respectively. Solid red circles and solid green squares represent the FM and PM with offset in exchange energy (Eq.\ref{eq:Ham_mag_offset}), respectively. Vertical dash lines show the corresponding equilibrium volumes for each case.  (Right) Spontaneous volume magnetostriction ($\omega_s$) obtained with -mode 3 of MAELAS using a second order (open blue circles) and third order (open green squares) polynomial fit (Eq.\ref{eq:dE_tot_fit}), where the volume of the used reference state is  $V_{PM}=V^{\text{offset}}_{FM}=11.7767148$ \r{A}$^3$/atom  (limit at infinite supercell size for the PM state), and the energy of the deformed unit cells was evaluated without offset in the exchange energy. Horizontal red dash line shows the calculated values with the atomistic simulations without offset in the exchange energy, while pink dot line corresponds to the atomistic case with offset in exchange energy (Eq.\ref{eq:Ham_mag_offset}). Orange line stands for the theoretical value given by Eq.\ref{eq:ws_theory} evaluated at volume $V_{PM}=V^{\text{offset}}_{FM}=11.7767148$ \r{A}$^3$/atom.}
\label{fig:lmp}
\end{figure}


\subsubsection{Atomistic simulations}
\label{subsection:atom_lmp}

Firstly, we compute the equation of state for the FM and PM states. We use  a bcc structure with cell size 1x1x1 (2 atoms/cell) for the FM state and 30x30x30 (54000 atoms/cell) for the PM state, respectively. The evaluation of the energy is performed using the SPIN package\cite{TRANCHIDA2018406} of LAMMPS\cite{thompson2022lammps}. The random distribution of the spins for the PM state is achieved with the implemented random generator of the SPIN package\cite{TRANCHIDA2018406} of LAMMPS\cite{thompson2022lammps} with seed number 31. The equilibrium volume for the FM and PM states ($V_{FM}=11.6592907$ \r{A}$^3$/atom and $V_{PM}=11.7758592$ \r{A}$^3$/atom) are derived from a fitting to the Murnaghan EOS\cite{Murnaghan1944,Fu} leading to $\omega^{\text{atomistic}}_s=-0.009899$, see Fig.\ref{fig:lmp}. It is instructive to calculate these equilibrium volumes using the following offset in the exchange energy
\begin{equation}
\begin{aligned}
\mathcal{H}^{\text{offset}}_{mag}(\boldsymbol{r},\boldsymbol{s}) & =  -\frac{1}{2}\sum_{i,j=1,i\neq j}^N J(r_{ij})[\boldsymbol{s}_i\cdot\boldsymbol{s}_j-1].
\label{eq:Ham_mag_offset}
\end{aligned}
\end{equation}
This offset is typically included in classical spin-lattice models to ensure that the pressure is zero at the minimum energy  \cite{nieves2021spinlattice_prb}. In this case, we find $V^{\text{offset}}_{FM}=11.7767148$ \r{A}$^3$/atom and $V^{\text{offset}}_{PM}=11.8959898$ \r{A}$^3$/atom, which leads to approximately the same value for $\omega^{\text{atomistic(offset)}}_s=-0.01002$. This small difference with respect to the case without offset may be related to finite size effects arising from the supercell size used in the simulation of the PM state. Note that the EOS of the PM state without offset is the same as the FM state with offset since in both cases the exchange energy is zero, see Fig.\ref{fig:lmp}. Consequently, the equilibrium volume of the PM state without offset (in the limit at infinite supercell size) is the same as the equilibrium volume of the FM with offset ($V_{PM}=V^{\text{offset}}_{FM}=11.7767148$ \r{A}$^3$/atom). We point out that this fact may cause misunderstanding about the volume of the reference state (unstrained state) required in -mode 3 of MAELAS.

\subsubsection{MAELAS}
\label{subsection:maelas_lmp}

 As described in Section \ref{subsection:method_iso}, one needs to know $V_{PM}$ in order to build the reference state required in -mode 3 of MAELAS to compute $\omega_s=(V_{FM}-V_{PM})/V_{PM}$. Here, we use the value obtained with the atomistic simulation without offset (in the limit at infinite supercell size for the PM state) $V_{PM}=V^{\text{offset}}_{FM}=11.7767148$ \r{A}$^3$/atom. In Fig.\ref{fig:lmp} we  show the calculation of $\omega_s$  with -mode 3 of MAELAS using a second order and third order  polynomial fit (Eq.\ref{eq:dE_tot_fit}). We observe a good agreement between MAELAS and atomsitic results, where the third order  polynomial fit is more accurate and robust than the second order in a broad range of maximum applied strains ($s_{max}\in[0.007,0.018]$). The same result for $\omega_s$ is found using the PM volume including the offset $V^{\text{offset}}_{PM}=11.8959898$ \r{A}$^3$/atom to build the reference state for MAELAS, where in this case the evaluation of the exchange energy for each deformed unit cell should include the offset in Eq.\ref{eq:Ham_mag_offset}.  The source files of this test are available in Example folder \cite{maelasexample}. Here, we make use of the program Atomsk\cite{HIREL2015212} to convert input files to LAMMPS format\cite{thompson2022lammps,TRANCHIDA2018406}.

\subsubsection{Theory}
\label{subsection:theory_lmp}

We can further verify the above results by comparing them to the exact solution of $\omega_s$ for this spin-lattice model. Extending the derivation described by Chikazumi in Ref.\cite{Chika} up to second nearest neighbors for a bcc structure, we find 
\begin{equation}
\begin{aligned}
(bcc):\quad \omega_s^{\text{theory}}=\pm\omega_s^{J_1}\pm\omega_s^{J_2},
\label{eq:ws_theory}
\end{aligned}
\end{equation}
where
\begin{equation}
\begin{aligned}
\omega_s^{J_1} & = \frac{4}{V_{PM}(C_{11}+2C_{12})}r_{0,1}\frac{dJ}{dr}\Bigg\vert_{r=r_{0,1}} ,\\
\omega_s^{J_2} & = \frac{3}{V_{PM}(C_{11}+2C_{12})}r_{0,2}\frac{dJ}{dr}\Bigg\vert_{r=r_{0,2}},
\label{eq:ws_theory_J}
\end{aligned}
\end{equation}
where $V_{PM}$ is the equilibrium volume of the PM state in units of volume per atom ($[V_{PM}]=$\r{A}$^3$/atom), $r_{0,1}$ and $r_{0,2}$ are the interatomic distances between first and second nearest neighbors at equilibrium PM volume, respectively, and $dJ/dr$ is the spatial derivative of the exchange parameter in Eq.\ref{eq:BS_J}. The sign $+$ in Eq.\ref{eq:ws_theory} corresponds to parallel spin alignment between each neighbor shell at the GS, while the sign $-$ is for the case of anti-parallel alignment. The elastic constants are evaluated at $V_{PM}$ using the same magnetic order of the GS, finding $C_{11}=238.86$ GPa and $C_{12}=148.33$ GPa. Inserting the corresponding values of this spin-lattice model in Eq.\ref{eq:ws_theory} at $V_{PM}=V^{\text{offset}}_{FM}=11.7767148$ \r{A}$^3$/atom (limit at infinite supercell size for the PM state), we obtain $\omega^{\text{theory}}_s=-0.010105$ which is in good agreement with MAELAS and atomistic simulations, see Fig.\ref{fig:lmp}. This theoretical expression shows that the negative value of $\omega_s$ found in this model is due to  the negative spatial derivative of the exchange parameters ($dJ/dr<0$). The corresponding formula for $b^{\alpha,2}$ is found by substituting Eq.\ref{eq:ws_theory} in Eq.\ref{eq:ws_cub_total_iso_ani}.

\subsubsection{Possible extensions of the spin-lattice model}
\label{subsection:model_lmp}

More advanced spin-lattice models could include different Bethe-Slater curves for each nearest neighbor pair exchange interaction ($J_1(r)$, $J_2(r)$, ...) which could give a more realistic description of the spatial derivative of the exchange parameters \cite{Wang_J_r}. Moreover, an additional contribution to $\omega_s$ may arise from the volume dependence of magnetic moments (relevant in itinerant magnets) through the Landau energy, which could be included in the model via the Heisenberg-Landau Hamiltonian \cite{Ma2012}.

\subsection{First principles calculations}
\label{section:dft}

Now, we show some examples of this method using first principles calculations. To this end, MAELAS is interfaced with VASP \cite{vasp_1,vasp_2,vasp_3} to compute the energies of each deformed state. We consider the same materials (bcc Fe, fcc Ni, hcp Co, Fe$_2$Si space group (SG) 164, L1$_0$ FePd and YCo SG 63) as in the publication of the previous versions of MAELAS, where the anisotropic quantities were studied \cite{maelas_publication2021,maelas_v2}. Unlike the classical spin-lattice model, the calculation of $V_{PM}$ with Density Functional Theory (DFT) is not easy so that here we use reference states for -mode 3 of MAELAS with volume $V$ that may not be equal to $V_{PM}$. Hence, the presented values in this section should not be compared with theoretical and experimental data available in literature.

\subsubsection{bcc Fe}
\label{section:bcc_Fe}

Let us first apply this method to bcc Fe using spin-polarized DFT calculation (without SOC) of the energy, as described in Section \ref{subsection:method_iso}. We consider the conventional cubic unit cell of bcc Fe (2 atoms/cell). The interactions were described by the projector augmented wave (PAW) method \cite{vasp_4} potential with 14 valence electrons for bcc Fe, where we use the Generalized Gradient Approximation (GGA) with the Perdew-Burke-Ernzerhof (PBE) for the type of exchange-correlation \cite{Perdew} including  aspherical contributions to the PAW one-centre terms (LASPH = .TRUE.). The plane waves basis was generated for an energy cut-off of $586.476$ eV (two times larger than the default value). The energy convergence criterion of the electronic self-consistency was chosen as $10^{-7}$  eV/cell. Firstly, we calculate the equilibrium volume at the FM state ($V_{FM}$) by fitting the computed energy for different unit cell volume to the Vinet EOS\cite{vinet}. We obtain $V_{FM}=11.344498$ \r{A}$^3$/atom ($a^{FM}_0=2.83099$ \r{A}). This value is about $3.7\%$ smaller than the experimental value $V_{FM}^{exp}=11.77676$ \r{A}$^3$/atom of bcc Fe \cite{khol1967,Wang_J_r}. 

\begin{figure}[h!]
\centering
\includegraphics[width=\columnwidth ,angle=0]{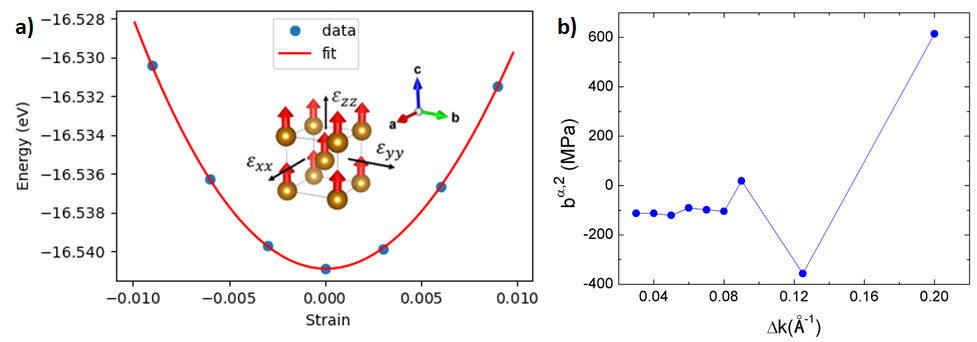}
\caption{(a) Fitting procedure to extract $b^{\alpha,2}$ for  bcc Fe. The unit cell  is deformed according to the parameterized strain tensor $\boldsymbol{\epsilon}(s)$ given in Table \ref{tab:method_data}. Blue circles give the DFT data, while the red solid line shows the fitting to a third order polynomial Eq.\ref{eq:dE_tot_fit}. The volume of the reference state is $V=11.344498$ \r{A}$^3$/atom, while the external pressure on this reference state is $P=0.063$ kB. (b) Analysis of the convergence of $b^{\alpha,2}$ with respect to the smallest allowed spacing between k-points
($\Delta k$) in the mesh of the Brillouin zone.}
\label{fig:Febcc}
\end{figure}

\begin{figure}[h]
\centering
\includegraphics[width=1\columnwidth ,angle=0]{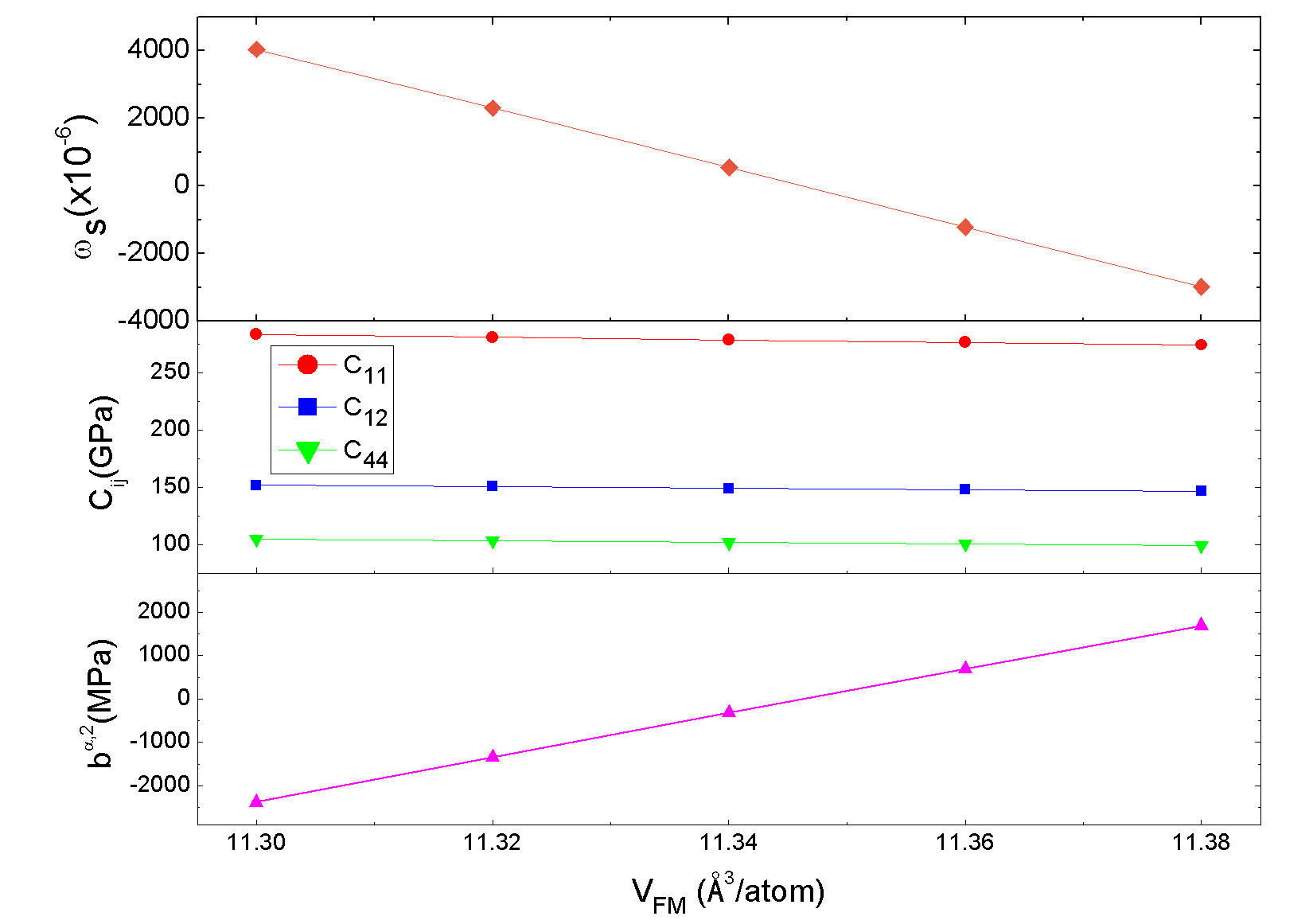}
\caption{Calculated isotropic magnetoelastic constant $b^{\alpha,2}$, elastic constants $C_{ij}$ and spontaneous volume magnetostriction $\omega_s$ for bcc Fe using different equilibrium volumes $V$ for the reference state.}
\label{fig:B_vol_Fe}
\end{figure}

\begin{figure}[h]
\centering
\includegraphics[width=0.74\columnwidth ,angle=0]{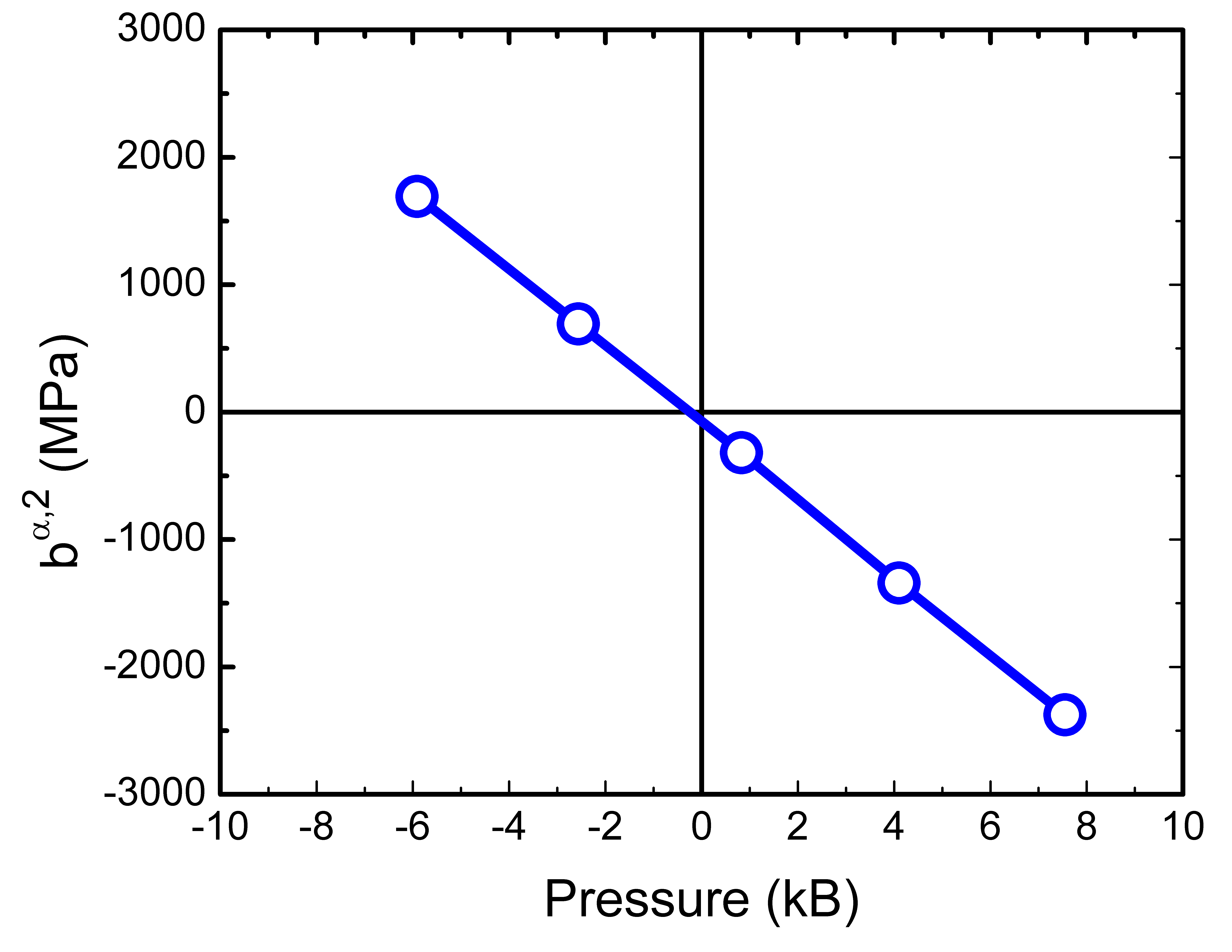}
\caption{Calculated isotropic magnetoelastic constant $b^{\alpha,2}$ for bcc Fe as a function of the external pressure on the reference state.}
\label{fig:b_pressure}
\end{figure}

Next, we use a reference state (unstrained state) with volume $V_{FM}$ for -mode 3 of MAELAS. The fitting procedure to extract $b^{\alpha,2}$ is shown in Fig.\ref{fig:Febcc}. Namely, we applied the strain tensor given in Table \ref{tab:method_data} to the conventional unit cell of bcc Fe  with the previously calculated equilibrium volume. In this calculation, we used a small spacing between k-points ($\Delta k=0.03$ \r{A}$^{-1}$) where we verified the convergence of $b^{\alpha,2}$ with respect to $\Delta k$ (input parameter called KSPACING in VASP), see Fig.\ref{fig:Febcc}. From the third order polynomial fitting (Eq.\ref{eq:dE_tot_fit}) of the energy versus strain (free parameter $s$) data, we obtain $b^{\alpha,2}=-112.5$ MPa. Similarly, using the same lattice parameter and VASP settings in the program AELAS\cite{AELAS}, we get the following elastic constants $C_{11}=279.6$ GPa, $C_{12}=149.1$ GPa and $C_{44}=101.5$ GPa, which are in reasonable good agreement with experiment\cite{Fe_elas_exp} and previous DFT calculations\cite{deJong2015,Mat_Proj_1,Fe_MP}. Inserting our calculated elastic and magnetoelastic constants in Eq.\ref{eq:ws_cub_total} gives $\omega^{MAELAS}_s(V_{FM})=1.95\times10^{-4}$. This small but finite value may be related to a low external pressure ($P=0.063$ kB) of the used reference state since MAELAS should give null exchange magnetostriction if the GS (zero external pressure $P=0$) is considered as the reference state ($b^{\alpha,2}_{MAELAS}(V_{FM})=\omega^{MAELAS}_s(V_{FM})=0$), see Section \ref{subsection:method_iso}. This fact is verified in Figs. \ref{fig:B_vol_Fe} and \ref{fig:b_pressure}, where we repeated this calculation using different volumes $V$ close to $V_{FM}$ for the reference state. The observed linear dependence of $\omega^{MAELAS}_s(V)$ on $V$ follows Eq.\ref{eq:ws_maelas_gs_taylor}.

Unfortunately, the value of $\omega_s=(V_{FM}-V_{PM})/V_{PM}$ cannot be estimated from these results since $V_{PM}$ is unknown. An experimental value measured by Richter and  Lotter\cite{richter} is $\omega_s=(4\pm0.4)\times10^{-4}$, which  one order of magnitude smaller than those  reported by Ridley and Stuart\cite{Ridley_1968} $\omega_s=3.3\times10^{-3}$ and $\omega_s=7.5\times10^{-3}$. Experimental negative values have been also reported ($\omega_s=-2.7\times10^{-3}$\cite{booktremolet} and $\omega_s=-3.6\times10^{-3}$\cite{Carr}), which might be consistent with the large negative spatial derivative of the first nearest neighbor exchange parameter ($dJ_1/dr<0$) found with DFT calculations\cite{Wang_J_r}, see Eq.\ref{eq:ws_theory}. A theoretical value about $\omega_s\sim 10^{-3}$ was calculated by Khmelevskyi and Mohn  using  first-principles calculations with the DLM\cite{khmelevskyi}, while higher values ($\omega_s> 10^{-2}$) have been reported within the Stoner model\cite{shimizu1978,kake1986}. 

\subsubsection{fcc Ni}

In the case of fcc Ni, we perform spin-polarized calculations without SOC based on the method described in Section \ref{subsection:method_iso}. We consider the conventional  unit cell (4 atoms/cell). The interactions were described by PAW method \cite{vasp_4} potential with 16 valence electrons, where we use GGA with PBE for the type of exchange-correlation \cite{Perdew}. The plane waves basis was generated for an energy cut-off of $735.972$ eV (two times larger than the default value). The energy convergence criterion of the electronic self-consistency was chosen as $10^{-7}$  eV/cell. Firstly, we calculate the equilibrium volume at the FM state ($V_{FM}$) by fitting the computed energy for different unit cell volume to the Vinet EOS\cite{vinet}. We obtained the equilibrium volume $V_{FM}=10.90803$ \r{A}$^3$/atom ($a_0^{FM}=3.52048$ \r{A}). This value is about $0.3\%$ greater than the experimental value $V_{FM}^{exp}=10.8797$ \r{A}$^3$/atom \cite{MORUZZI197811}.

\begin{figure}[h!]
\centering
\includegraphics[width=\columnwidth ,angle=0]{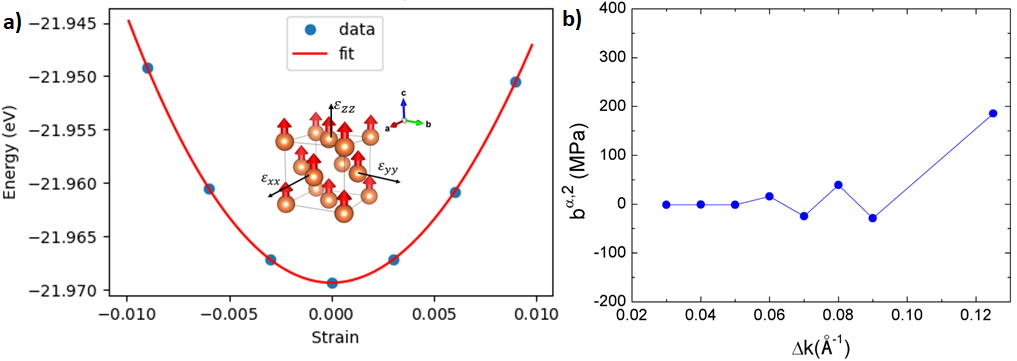}
\caption{(a) Fitting procedure to extract $b^{\alpha,2}$ for  fcc Ni. The unit cell  is deformed according to the parameterized strain tensor $\boldsymbol{\epsilon}(s)$ given in Table \ref{tab:method_data}. Blue circles give the DFT data, while the red solid line shows the fitting to a third order polynomial Eq.\ref{eq:dE_tot_fit}. (b) Analysis of the convergence of $b^{\alpha,2}$ with respect to the smallest allowed spacing between k-points
($\Delta k$) in the mesh of the Brillouin zone.}
\label{fig:Nifcc}
\end{figure}

\begin{figure}[h]
\centering
\includegraphics[width=1\columnwidth ,angle=0]{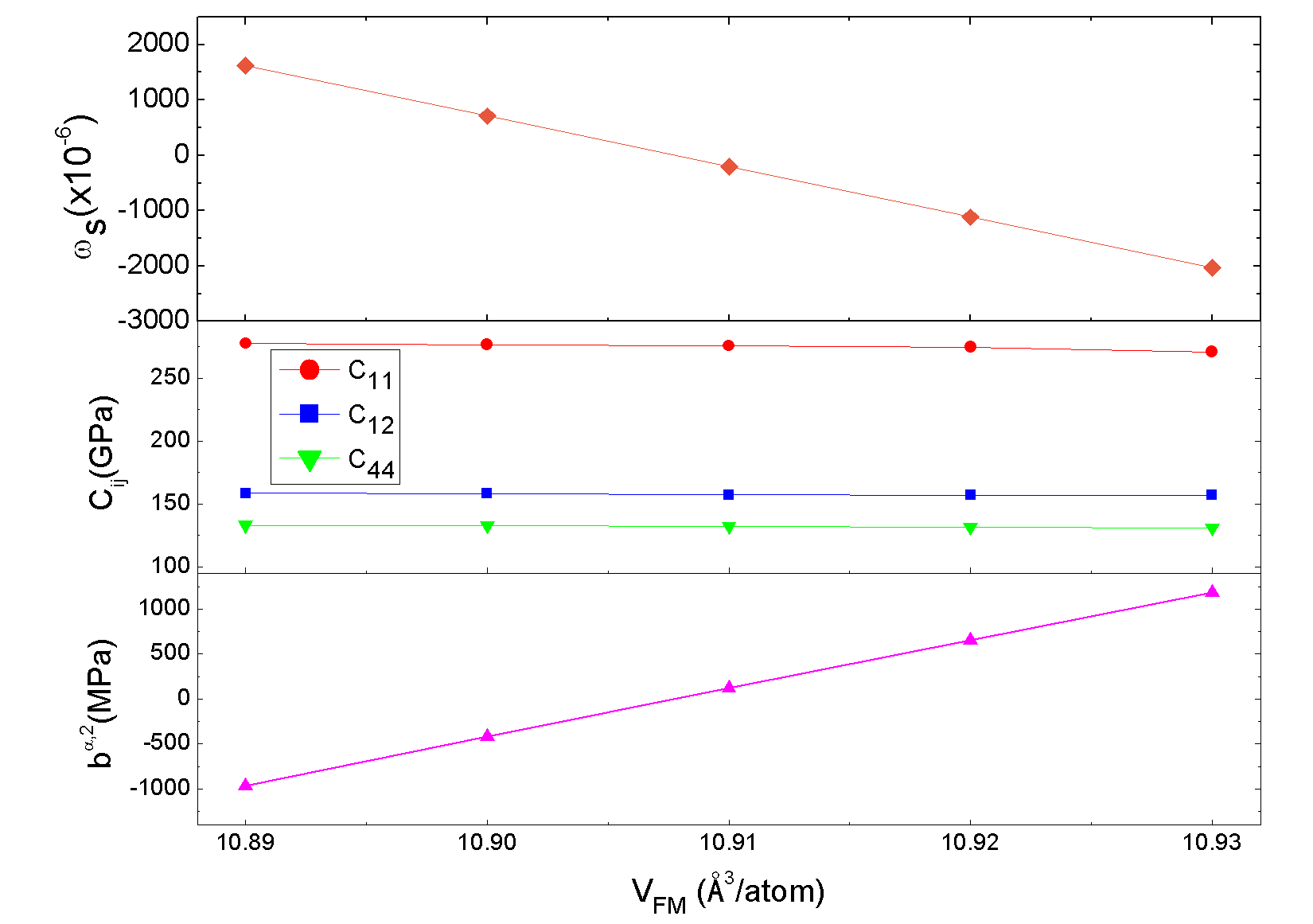}
\caption{Calculated isotropic magnetoelastic constant $b^{\alpha,2}$, elastic constants $C_{ij}$ and spontaneous volume magnetostriction $\omega_s$ for fcc Ni using different equilibrium volumes $V$ of the reference state.}
\label{fig:B_vol_Ni}
\end{figure}

The fitting procedure to extract $b^{\alpha,2}$ is shown in Fig.\ref{fig:Nifcc}. From the third order polynomial fitting  (Eq.\ref{eq:dE_tot_fit}) of the energy versus strain data, we obtain $b^{\alpha,2}=-1.3$ MPa. Similarly, using the same lattice parameter and VASP settings in the program AELAS\cite{AELAS}, we get the following elastic constants $C_{11}=275.9$ GPa, $C_{12}=157.5$ GPa and $C_{44}=132.2$ GPa which are in reasonable good agreement with experiment\cite{Ni_elas_exp} and previous DFT calculations\cite{deJong2015,Mat_Proj_1,Ni_MP}. Inserting our calculated elastic and magnetoelastic constants in Eq.\ref{eq:ws_cub_total} gives $\omega_s=2.3\times10^{-6}$. This negligible exchange magnetostriction is related to the fact that the reference state is very close to the GS, see Section \ref{subsection:method_iso}. In Fig.\ref{fig:B_vol_Ni}, we repeated this calculation using different volumes $V$ close to $V_{FM}$ for the reference state, where the linear behaviour of $\omega^{MAELAS}_s(V)$ is described by Eq.\ref{eq:ws_maelas_gs_taylor}.

As in the case of bcc Fe, the value of $V_{PM}$ is unknown so we cannot estimate $\omega_s$ for this material. The experimental value found by Williams\cite{williams} is $\omega_s=(3.24\pm0.15)\times10^{-4}$ and the theoretical result calculated by Shimizu\cite{shimizu1978} is $\omega_s=3.75\times10^{-4}$ using the Stoner model. Larger theoretical values than these ones have been also reported ($\omega_s>10^{-3}$)\cite{Janak1976,kake1986}. We point out that negative values of $\omega_s$ have also been reported  for fcc Ni\cite{richter,tanji,kake1986,booktremolet,Carr}.

\begin{table}[h!]
\centering
\caption{Isotropic magnetoelastic constants, isotropic magnetostrictive coefficients and isotropic contribution to spontaneous volume magnetostriction calculated with the new method (-mode 3) available in MAELAS v3.0. The symbol [SOC] means that this quantity has been calculated with SOC following the methodology described in Section \ref{subsection:method_ani}, where the anisotropic magnetoelastic constants (required to compute $\Lambda(b^{ani})$) are taken from  the publication of version 2.0 \cite{maelas_v2} using -mode 2. The quantities without symbol [SOC]  were calculated without SOC according to the methodology described in Section \ref{subsection:method_iso}. Note that these calculations depend strongly on the volume $V$ of the reference state, which is shown in the first column. The spontaneous volume magnetostriction given by MAELAS corresponds to $\omega^{MAELAS}_s(V)=(V_{GS}-V)/V$.  The used volume $V$ in these calculations may not be equal to $V_{PM}$, so that the presented values should not be compared with theoretical and experimental data for $\omega_s=(V_{GS}-V_{PM})/V_{PM}$ available in literature.}
\label{tab:tests_MAELAS}
\resizebox{\textwidth}{!}{%
\begin{tabular}{@{}ccc ccc ccc@{}}
\toprule
Material &
  Crystal system &
  \begin{tabular}[c]{@{}c@{}} DFT \\ Exchange  \\ Correlation\end{tabular}  &
  $b^{iso}$(V) &
  \begin{tabular}[c]{@{}c@{}}MAELAS\\-mode 3\\  (MPa)\end{tabular}  &
  $\lambda^{iso}$(V) &
  \begin{tabular}[c]{@{}c@{}}MAELAS\\-mode 3\\ ($\times10^{-6}$)\end{tabular} & 
  $\omega^{MAELAS}_s(V)$ &
  \begin{tabular}[c]{@{}c@{}}MAELAS\\-mode 3\\  ($\times10^{-6}$)\end{tabular} \\ \midrule \hline
bcc Fe      & Cubic (I)      &  GGA & $b^{\alpha,2}$        & -112.5  & $\lambda^{\alpha,2}$   & 194.7   & $\omega_s^{iso}$ & 194.7  \\
      $a_0=2.83099$\r{A}      &       SG 229         &     &        &  &  &    &  &    \\ 
      $V=11.344498$ \r{A}$^3$/atom      &             &     &        &  &  &    &  &    \\
      $P=0.063$ kB     &             &     &        &  &  &    &  &    \\\midrule
fcc Ni      & Cubic (I)      &  GGA & $b^{\alpha,2}$        & -1.3  & $\lambda^{\alpha,2}$   & 2.3  & $\omega_s^{iso}$ & 2.3  \\
   $a_0=3.52048$\r{A}         &       SG 225         &     &         &  &  &  &  &  \\ 
    $V=10.90803$ \r{A}$^3$/atom        &               &     &         &  &  &  &  &  \\
       $P=-1.003$ kB       &                &     &         &  &  &  &  &  \\\midrule
hcp Co      & Hexagonal (I)  &  LSDA+U  & $b_{11}$  & -316.5   & $\lambda^{\alpha1,0}$ & -2260.8 & $\omega_s^{iso}$ &  6335.9  \\
   $a_0=2.48896$\r{A}       &          SG 194      &  $J=0.8$eV    & $b_{12}$  & -2755.9   & $\lambda^{\alpha2,0}$ & 10857.6  & & \\
    $c_0=4.02347$\r{A}       &                &  $U=3$eV    & $b_{11}$[SOC]  & -383.2   & $\lambda^{\alpha1,0}$[SOC] & -2159.3 & $\omega_s^{iso}$[SOC] &  6675.2   \\
      $V=10.792901$ \r{A}$^3$/atom      &                &     & $b_{12}$[SOC]  & -2824.2  & $\lambda^{\alpha2,0}$[SOC] & 10993.9 &  &    \\
      $P=-5.27$ kB      &                &     &   &  &  &  &  &    \\
            \midrule
            Fe$_2$Si     & Trigonal (I)  &  GGA  & $b_{11}$  & -94.1   & $\lambda^{\alpha1,0}$ & 134.1 & $\omega_s^{iso}$ &  378.5   \\
   $a_0=3.9249$\r{A}       &          SG 164      &     & $b_{12}$  & -83.6  & $\lambda^{\alpha2,0}$ & 110.2  & &  \\
    $c_0=4.8311$\r{A}       &                &     & $b_{11}$[SOC]  & -135.4   & $\lambda^{\alpha1,0}$[SOC] & 191.2 & $\omega_s^{iso}$[SOC] &  549.1  \\
     $V=10.741922$ \r{A}$^3$/atom        &                &     & $b_{12}$[SOC]  & -123.3  & $\lambda^{\alpha2,0}$[SOC] & 166.7 &  &   \\
       $P=-0.4$ kB   &                &     &   &   &  & &  &   \\
            \midrule    
            L1$_0$ FePd & Tetragonal (I) & GGA  & $b_{11}$  & -2123.9   & $\lambda^{\alpha1,0}$ & 4347.2 & $\omega_s^{iso}$ &  11852.5  \\
   $a_0=2.6973$\r{A}       &          SG 123      &     & $b_{12}$  & -1993.3  & $\lambda^{\alpha2,0}$ & 3158.2  & &  \\
    $c_0=3.7593$\r{A}       &                &     & $b_{11}$[SOC]  & -2086.4   & $\lambda^{\alpha1,0}$[SOC] & 4179.3 & $\omega_s^{iso}$[SOC] & 11728.7     \\
       $V=13.675852$ \r{A}$^3$/atom     &                &     & $b_{12}$[SOC]  & -2004.3 & $\lambda^{\alpha2,0}$[SOC] & 3370.0 &  &    \\
       $P=12.83$ kB   &                &     &   &   &  & &  &   \\
            \midrule 
            YCo & Orthorhombic &  LSDA+U &   $b_{1}^{\alpha,0}$  & -522.3   & $\lambda_1^{\alpha,0}$ & 2258.8 & $\omega_s^{iso}$ &  2258.8  \\
   $a_0=4.0686$\r{A}       &          SG 63      &  $J=0.8$eV    & $b_{2}^{\alpha,0}$  & 14.7  & $\lambda_2^{\alpha,0}$ & -946.6 & &  \\
    $b_0=10.3157$\r{A}       &                &   $U=1.9$eV   & $b_{3}^{\alpha,0}$  & 85.1   & $\lambda_3^{\alpha,0}$ & -1171.9 &  &   \\
     $c_0=3.8957$\r{A}       &                &     & $b_{1}^{\alpha,0}$[SOC]  & -512.8 & $\lambda_1^{\alpha,0}$[SOC] & 2148.6 &  $\omega_s^{iso}$[SOC] & 2148.6    \\
        $V=20.438039$ \r{A}$^3$/atom     &                &     & $b_{2}^{\alpha,0}$[SOC]  & -38.1 & $\lambda_2^{\alpha,0}$[SOC] & -274.6 &  &     \\
        $P=-0.64$ kB      &                &     & $b_{3}^{\alpha,0}$[SOC]  & 81.3 & $\lambda_3^{\alpha,0}$[SOC] & -1019.8 &  &     \\
            \bottomrule
\end{tabular}%
}
\end{table}

\subsubsection{Symmetries lower than cubic}

\begin{figure}[h]
\centering
\includegraphics[width=0.7\columnwidth ,angle=0]{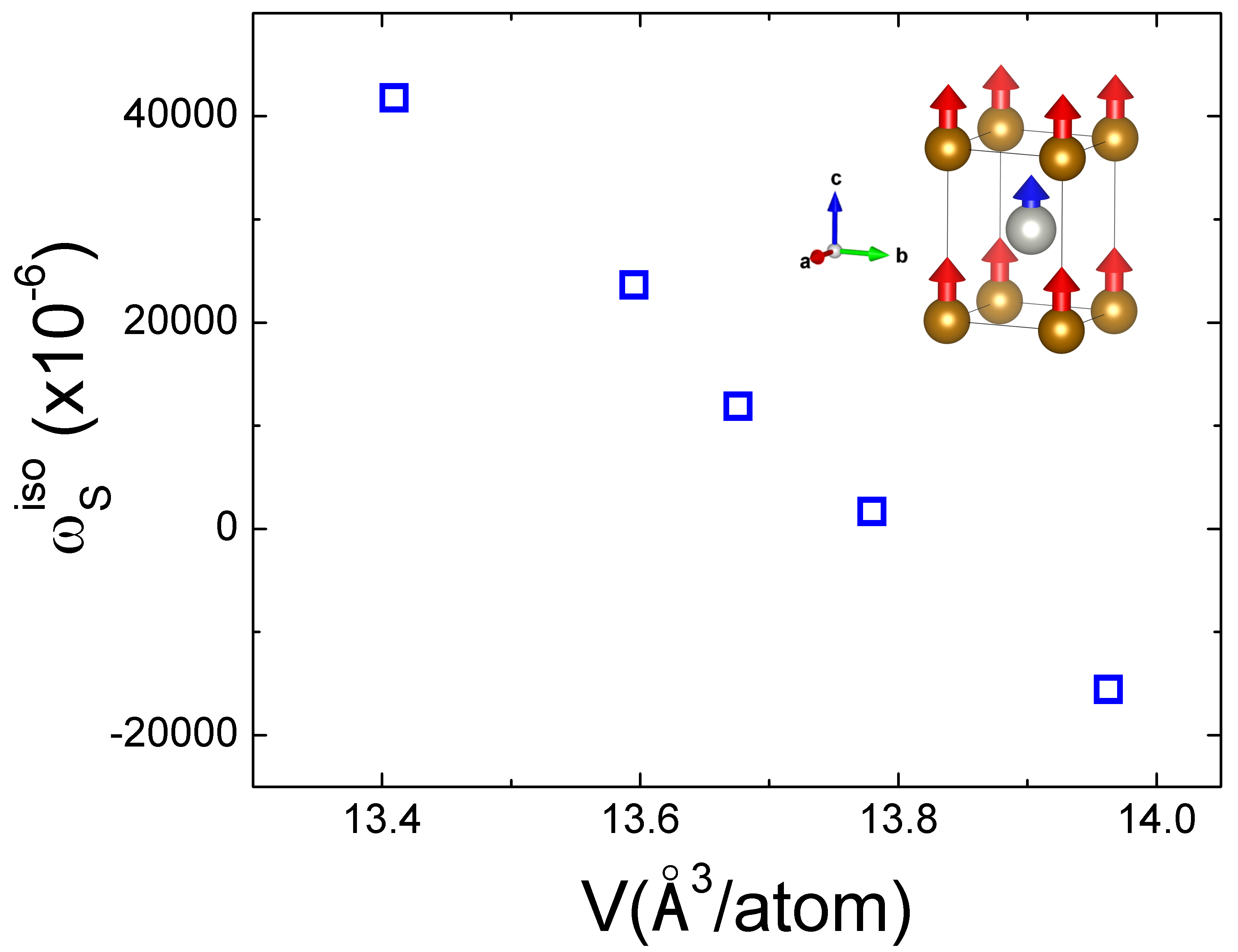}
\caption{Calculated isotropic contribution to spontaneous volume magnetostriction $\omega^{iso}_s$ for tetragonal L1$_0$ FePd using different equilibrium volumes $V$ of the reference state.}
\label{fig:FePd_ws}
\end{figure}

As an example of materials with symmetry lower than cubic, we consider the same materials (hcp Co, Fe$_2$Si space group (SG) 164, L1$_0$ FePd and YCo SG 63) as in the publication of previous versions of MAELAS \cite{maelas_publication2021,maelas_v2}, and we use exactly the same relaxed unit cells (volume of the reference state), VASP settings (energy cut-off, pseudopotentials, exchange correlation,...) and elastic constants calculated with AELAS \cite{maelas_publication2021,AELAS}. We computed $b^{iso}$, $\lambda^{iso}$ and $\omega_s^{iso}$ with the two approaches described in Section \ref{section:method}, it means via spin-polarized calculations without SOC (Section \ref{subsection:method_iso}) and with SOC (Section \ref{subsection:method_ani}), where in the latter the anisotropic magnetoelastic constants (required to compute $\Lambda(b^{ani})$) are taken from  the publication of version 2.0 \cite{maelas_v2} using -mode 2. The obtained results for each material are shown in Table \ref{tab:tests_MAELAS}. In general, we see similar values using these two methods, except for $b_{2}^{\alpha,0}$ of YCo where the sign changed after including SOC. Inconsistent results between these two approaches could be related to numerical reasons (accuracy in the evaluation of energy, magnetoelastic constants, computational error in the DFT calculation, etc.) and physical reasons like strong SOC effect on the equilibrium volume of the GS. Note that the used volume $V$ may not be equal to $V_{PM}$, so that the presented values in Table \ref{tab:tests_MAELAS} should not be compared with theoretical and experimental data available in literature for $\omega_s=(V_{FM}-V_{PM})/V_{PM}$. In Fig.\ref{fig:FePd_ws}, as an example of uniaxial crystal system, we verified the linear dependence of $\omega_s^{MAELAS}$ on the volume of the reference state for tetragonal L1$_0$ FePd described by Eq.\ref{eq:ws_maelas_gs_taylor}. Here, we took into account the correct expression for $\lambda^{\alpha 1,0}$ given by Eq.\ref{eq:lamb_hex_iso_correct} which was incorrectly written in previous publications \cite{CLARK1980531,Cullen,maelas_publication2021}. Similarly, we also verified this linear behaviour in orthorhombic crystal YCo.

\section{Limitations and future perspectives}
\label{section:limit}

The main limitations of the new presented methodology arise from the terms not included in the elastic and magnetoelastic energy in Eq.\ref{eq:E_tot} or Eq.\ref{eq:E_tot_ani}. In this method, the elastic energy is considered up to second order in the strain, and the magnetoelastic energy contains only linear terms in the strain up to second order in the magnetization direction $\alpha$. For instance, the terms in the elastic energy in third order in the strain jointly with the terms in the magnetoelastic energy in the second order in the strain are responsible of morphic effects on sound velocity \cite{DUTREMOLETDELACHEISSERIE198277,nieves_sound2022,booktremolet}. Possible improvements of current version of MAELAS could be the addition of missing crystal symmetries: Cubic (II), Hexagonal (II), Trigonal (II), Tetragonal (II), Monoclinic and Triclinic.

The presented methodology is based on total energy, treating the volume as independent variable. The development of alternative methods that can complement and improve this methodology is an interesting topic to explore in the future. For example, one could try to consider the volume, the magnetic moments, and the magnetostriction as functions of an external pressure. Such an approach, however, should be based on a different thermodynamic potential (the enthalpy) instead of the total energy.

\section{Conclusions}
\label{section:con}

The presented methodology based on the magnetoelastic energy provides a simple way to compute the isotropic magnetoelastic constants, isotropic magnetostrictive coefficients and spontaneous volume magnetostriction arising from the exchange magnetostriction. As input, this approach requires a reference state that combines the equilibrium volume of the PM state and magnetic order of the GS. Unfortunately, the difficulties in determination of the equilibrium volume of the PM state with first-principles calculations may limit the range of applicability of this method like for high-throughput screening techniques. We also found that if the GS is used as reference state, then a null exchange magnetostriction is obtained. The theoretical analysis of the classical spin-lattice model shows that the spatial derivative of exchange parameters plays an important role in exchange magnetostriction.

\section*{Declaration of competing interest}

The authors declare that they have no known competing financial interests or personal relationships that could have appeared to influence the work reported in this paper.

\section*{Data availability}

Information about the MAELAS package is available on the
GitHub repository \cite{Maelas}.

\section*{Acknowledgement}

This work was supported by the ERDF in the IT4Innovations national supercomputing center - path to exascale project (CZ.02.1.01/0.0/0.0/16-013/0001791) within the OPRDE and projects “e-INFRA CZ (ID:90140)" and Donau No. 8X20050 by The Ministry of Education, Youth and Sports of the Czech Republic. The work was supported also by Czech Science Foundation of the Czech Republic by grant No. 22-35410K. This work was supported by the National Key Research and Development Program of China (No. 2017YFB0702100), National Thousand Young Talents Program of China, the Fundamental Research Funds for the Central Universities.

\appendix

\section{Analysis of the isotropic term of the magnetoelastic energy}
\label{app_matrix}

Unlike the theory of elasticity where there is a great consensus on the definitions of the elastic constants, in the theory of magnetoelasticity one can find a wide variety of different conventions for the magnetoelastic constants \cite{maelas_publication2021}. In some of these conventions, the contribution from isotropic and anisotropic magnetic interactions are mixed in the definitions of magnetoelastic constants. In 1993, E. du Tremolet de Lacheisserie proposed an universal notation (Lacheisserie convention) based on the framework of group theory which is valid for any crystal symmetry \cite{booktremolet}. The Lacheisserie convention has also the advantage that fully decouples the contribution from isotropic and anisotropic magnetic interactions to the magnetoelastic energy \cite{booktremolet}, which is the fundamental basis of the methodology implemented in -mode 3 of MAELAS. Below we analyze the isotropic term of the magnetoelastic energy ($E_{me}^{iso}$) using the Lacheisserie convention for the main crystal symmetries currently supported by MAELAS. The new version of MAELAS (v3.0) also prints the results using this universal notation (see Table \ref{tab:calculation_data}). 

\subsection{Cubic (I)}

In Lacheisserie convention, the magnetoelastic energy of cubic (I) crystals (point groups $432$, $\bar{4}3m$, $m\bar{3}m$) reads\cite{booktremolet}
\begin{equation}
    \begin{aligned}
      \frac{1}{V_0}  E_{me} & = \frac{1}{3}b^{\alpha,2}(\epsilon_{xx}+\epsilon_{yy}+\epsilon_{zz})\\
     & + b^{\gamma,2}\left(\left[\alpha_x^2-\frac{1}{3}\right]\epsilon_{xx}+\left[\alpha_y^2-\frac{1}{3}\right]\epsilon_{yy}+\left[\alpha_z^2-\frac{1}{3}\right]\epsilon_{zz}\right)\\
     & + 2 b^{\epsilon,2}(\alpha_{y}\alpha_{z}\epsilon_{yz}+\alpha_{z}\alpha_{x}\epsilon_{zx}+\alpha_{x}\alpha_{y}\epsilon_{xy}),
    \label{eq:Eme_universal}
    \end{aligned}
\end{equation}
where $V_0$ is the equilibrium volume, $\epsilon_{ij}$ is the strain tensor, $\alpha_{i}$ ($i=x,y,z$) are direction cosines of the magnetization, $b^{\alpha,2}$ is the isotropic magnetoelastic constant ($b^{\alpha,2}=b^{iso}$), and  $b^{\mu,2}$ ($\mu=\gamma,\epsilon$) are the anisotropic magnetoelastic constants ($b^{\mu,2}=b^{ani}$). The superscript greek letters ($\mu=\alpha,\gamma,\epsilon$) follow the Bethe’s group-theoretical notation of the irreducible representations \cite{booktremolet}. The definition of the magnetoelastic constants in Eq.\ref{eq:Eme_universal} has the advantage that fully decouples the isotropic and anisotropic magnetic interactions\cite{booktremolet}.  Namely, neglecting higher order corrections in Eq.\ref{eq:Eme_universal}, $b^{\alpha,2}$ contains all contribution to the magnetoelastic energy from isotropic magnetic interactions like the isotropic exchange, while $b^{\gamma,2}$ and $b^{\epsilon,2}$ contain all contribution to the magnetoelastic energy provided by anisotropic magnetic interactions like SOC, anisotropic exchange and crystal electric 
field interactions \cite{booktremolet}. Hence, within this convention we can split the magnetoelastic energy into its isotropic ($E_{me}^{iso}$) and anisotropic ($E_{me}^{ani}$) parts as  
\begin{equation}
    \begin{aligned}
      \frac{1}{V_0}  E_{me}  = \frac{1}{V_0}  (E_{me}^{iso}+E_{me}^{ani}),
    \label{eq:Eme_universal_2}
    \end{aligned}
\end{equation}
where
\begin{equation}
    \begin{aligned}
      \frac{1}{V_0}  E_{me}^{iso} & = \frac{1}{3}b^{\alpha,2}(\epsilon_{xx}+\epsilon_{yy}+\epsilon_{zz}),\\
     \frac{1}{V_0}  E_{me}^{ani} & = b^{\gamma,2}\left(\left[\alpha_x^2-\frac{1}{3}\right]\epsilon_{xx}+\left[\alpha_y^2-\frac{1}{3}\right]\epsilon_{yy}+\left[\alpha_z^2-\frac{1}{3}\right]\epsilon_{zz}\right)\\
     & + 2 b^{\epsilon,2}(\alpha_{y}\alpha_{z}\epsilon_{yz}+\alpha_{z}\alpha_{x}\epsilon_{zx}+\alpha_{x}\alpha_{y}\epsilon_{xy}).
    \label{eq:Eme_universal_3}
    \end{aligned}
\end{equation}
Here, one can verify that the form of $E^{ani}_{me}$ fully arises from the anisotropic magnetic interactions described effectively by the dipole term of the N\'{e}el model for a cubic crystal \cite{Chika}, decoupling isotropic and anisotropic magnetic interactions in the definition of the magnetoelastic constants. Alternatively, it is customary to write the magnetoelastic energy using Clark notation\cite{CLARK1980531}
\begin{equation}
    \begin{aligned}
       \frac{1}{V_0} E_{me} & =    b_0(\epsilon_{xx}+\epsilon_{yy}+\epsilon_{zz})+b_1(\alpha_x^2\epsilon_{xx}+\alpha_y^2\epsilon_{yy}+\alpha_z^2\epsilon_{zz})\\
    & +  2b_2(\alpha_x\alpha_y\epsilon_{xy}+\alpha_x\alpha_z\epsilon_{xz}+\alpha_y\alpha_z\epsilon_{yz}), 
    \label{eq:Eme_1}
    \end{aligned}
\end{equation}
where $b_i$ ($i=0,1,2$) are the magnetoelastic constants, which are related to Lacheisserie convention through the following equations
\begin{equation}
    \begin{aligned}
         b_0=\frac{1}{3}(b^{\alpha,2}-b^{\gamma,2}),\quad b_1=b^{\gamma,2},\quad b_2=b^{\epsilon,2}. 
    \label{eq:b0}
    \end{aligned}
\end{equation}
We see that the definition of $b_0$ mixes the contribution of isotropic and anisotropic magnetic interactions to the magnetoelastic energy, so that it cannot be calculated with a single mode implemented in MAELAS automatically. However, it  can be manually calculated by inserting
the isotropic magnetoelastic constant $b^{\alpha,2}$ (obtained with -mode 3) and the anisotropic one $b^{\gamma,2}$ (obtained with -mode 1 or -mode 2) in Eq.\ref{eq:b0}. 

\begin{figure}[h!]
\centering
\includegraphics[width=\columnwidth ,angle=0]{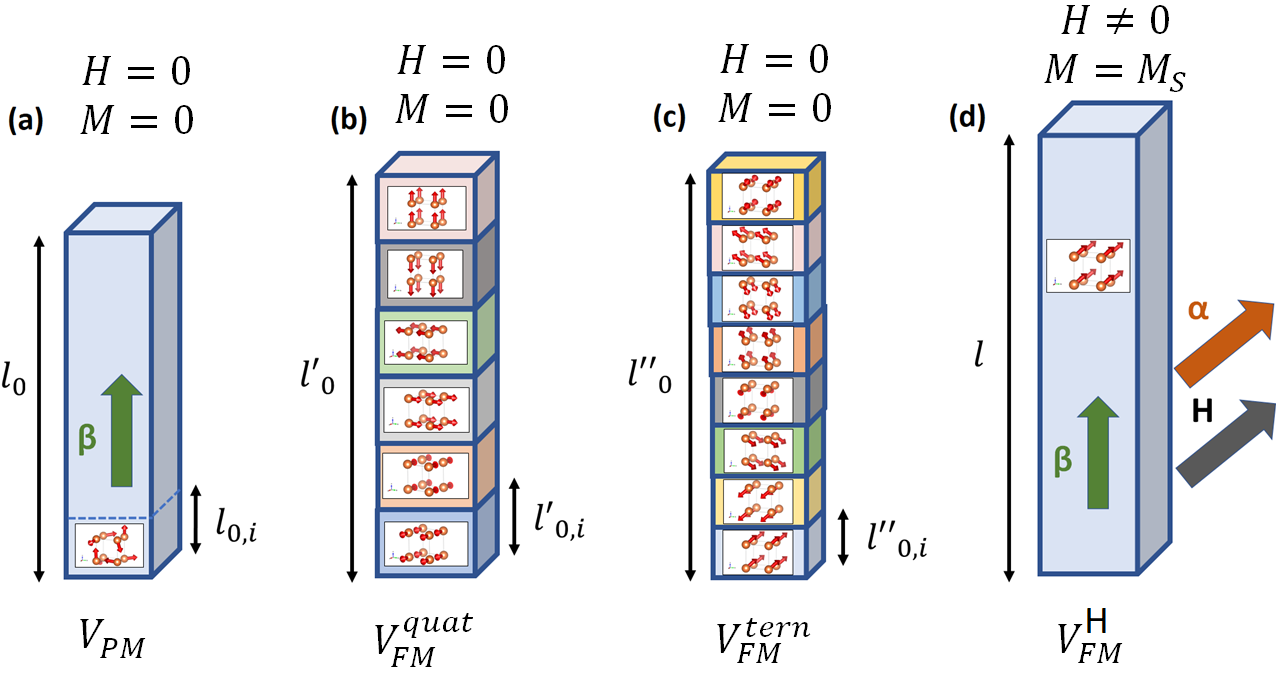}
\caption{(a) Ideal PM state of a cubic single crystal with randomly oriented atomic magnetic moments ($V_{PM}$). FM state with magnetic domains along all equivalent easy directions (b) quaternary ($V_{FM}^{quat}$, easy axis $<100>$) and (c) ternary ($V_{FM}^{tern}$, easy axis $<111>$). (d) Saturated magnetic state ($V_{FM}^H$). $M$, $M_s$ and $H$ represent the total magnetization, saturation magnetization and applied magnetic field, respectively.}
\label{fig:demag_cub}
\end{figure}

The Lacheisserie convention uses the same Bethe’s group-theoretical notation for the magnetostrictive coefficients. Let us as define $l_0$ and $l$ as the material lengths along the direction $\boldsymbol{\beta}$ when the system is at the PM and FM states, respectively. The fractional change in length $(l-l_0)/l_0$ can be obtained from the minimization of the elastic and magnetoelastic energies with respect to the strain\cite{CLARK1980531}. Equivalently, it can also be obtained through an expansion of cubic harmonic polynomials with respect to $\boldsymbol{\alpha}$ and normalized measuring length direction $\boldsymbol{\beta}$\cite{booktremolet,CLARK1980531}
\begin{equation}
\begin{aligned}
     \frac{l-l_0}{l_0}\Bigg\vert_{\boldsymbol{\beta}}^{\boldsymbol{\alpha}} & =\lambda^\alpha+\frac{3}{2}\lambda_{001}\left(\alpha_x^2\beta_{x}^2+\alpha_y^2\beta_{y}^2+\alpha_z^2\beta_{z}^2-\frac{1}{3}\right)\\
     & + 3\lambda_{111}(\alpha_x\alpha_y\beta_{x}\beta_{y}+\alpha_y\alpha_z\beta_{y}\beta_{z}+\alpha_x\alpha_z\beta_{x}\beta_{z})\\ & =\frac{1}{3}\lambda^{\alpha,2}+\lambda^{\gamma,2}\left(\alpha_x^2\beta_{x}^2+\alpha_y^2\beta_{y}^2+\alpha_z^2\beta_{z}^2-\frac{1}{3}\right)\\
     & + 2\lambda^{\epsilon,2}(\alpha_x\alpha_y\beta_{x}\beta_{y}+\alpha_y\alpha_z\beta_{y}\beta_{z}+\alpha_x\alpha_z\beta_{x}\beta_{z}),
    \label{eq:delta_l_cub_I}
\end{aligned}
\end{equation}
where $\lambda^{\alpha}=\lambda^{\alpha,2}/3$ is the isotropic magnetostrictive coefficients, and  $(3/2)\lambda_{001}=\lambda^{\gamma,2}$ and $3\lambda_{111}=2\lambda^{\epsilon,2}$ are the anisotropic magnetostrictive coefficients which are related to magnetoelastic and elastic constants via\cite{booktremolet}
\begin{equation}
    \begin{aligned}
        \lambda^{\alpha,2} & = -\frac{b^{\alpha,2}}{C_{11}+2C_{12}},\\
        \lambda^{\gamma,2} & = -\frac{b^{\gamma,2}}{C_{11}-C_{12}},\\
        \lambda^{\epsilon,2} & = -\frac{b^{\epsilon,2}}{C_{44}}.
    \label{eq:lamb_cub}
    \end{aligned}
\end{equation}
The spontaneous volume magnetostriction is defined as the fractional change in volume of the FM state ($V_{FM}$) with respect to the PM state ($V_{PM}$), that is\cite{khmelevskyi}
\begin{equation}
    \begin{aligned}
        \omega_s=\frac{V_{FM}-V_{PM}}{V_{PM}}.
    \label{eq:ws0}
    \end{aligned}
\end{equation}
This expression is widely used to compute $\omega_s$ through the direct calculation of $V_{FM}$ and $V_{PM}$ \cite{khmelevskyi,shimizu1978}. Let us write $V_{FM}=N_{cell}a^3$ and $V_{PM}=N_{cell}a_0^3$, where $N_{cell}$ is the number of unit cells in the system, $a$ and $a_0$ are the lattice parameter at FM and hypothetical PM states at temperature below the Curie temperature, respectively. Hence, Eq.\ref{eq:ws0} becomes\cite{ANDREEV199559}
\begin{equation}
    \begin{aligned}
        \omega_s=\frac{N_{cell}a^3-N_{cell}a_0^3}{N_{cell}a_0^3}=\left(\frac{a}{a_0}\right)^3-1=(\lambda_a+1)^3-1\simeq 3\lambda_a+\mathcal{O}(\lambda_a^2),
    \label{eq:ws00}
    \end{aligned}
\end{equation}
where $\lambda_a=(a-a_0)/a_0$ is the fractional change in length along the lattice vector $\boldsymbol{a}$, that is, $\boldsymbol{\beta}=(1,0,0)$. We see that $\omega_s$ depends on the definition of FM state. For example, if we consider a FM state where FM domains point to all equivalent easy directions (demagnetized state), then we have\cite{maelas_v2}
\begin{equation}
\begin{aligned}
     \lambda_a=\frac{l'_0-l_0}{l_0}\Bigg\vert_{\boldsymbol{\beta}=(1,0,0)}=\frac{l''_0-l_0}{l_0}\Bigg\vert_{\boldsymbol{\beta}=(1,0,0)} = \lambda^{\alpha},
    \label{eq:sum_delta_l_cub_I_4b}
\end{aligned}
\end{equation}
where $l'_0$ and $l''_0$ are the lengths at the FM state with easy quaternary (easy axis $<100>$, with volume $V_{FM}^{quat}$) and ternary (easy axis $<111>$, with volume $V_{FM}^{tern}$) directions, respectively, and $l_0$ is the length at the PM-like state (randomly oriented atomic magnetic moments). These magnetic states are schematically represented in Fig.\ref{fig:demag_cub}. Hence, replacing Eq.\ref{eq:sum_delta_l_cub_I_4b} in Eq.\ref{eq:ws00} gives
\begin{equation}
    \begin{aligned}
       \frac{V^{quat}_{FM}-V_{PM}}{V_{PM}}= \frac{V^{tern}_{FM}-V_{PM}}{V_{PM}} = \omega_s\simeq 3\lambda^\alpha=\lambda^{\alpha,2}=-\frac{b^{\alpha,2}}{C_{11}+2C_{12}}.
    \label{eq:ws000}
    \end{aligned}
\end{equation}
We observe that in this case the anisotropic interactions do not influence $\omega_s$.  Formally, we may decompose $\omega_s$ into isotropic and anisotropic contributions as
\begin{equation}
    \begin{aligned}
        \omega_s = \omega_s^{iso}+\omega_s^{ani},
    \label{eq:ws_cub_total}
    \end{aligned}
\end{equation}
where 
\begin{equation}
    \begin{aligned}
        \omega_s^{iso} & = 3\lambda^\alpha=\lambda^{\alpha,2}=-\frac{b^{\alpha,2}}{C_{11}+2C_{12}},\\
       \omega_s^{ani} & = 0.
    \label{eq:ws_cub_total_iso_ani}
    \end{aligned}
\end{equation}
The -mode 3 of MAELAS makes use of Eq.\ref{eq:ws_cub_total_iso_ani} to compute $\omega_s^{iso}$. 

It is important to note that the experimental measurement of these ideal volumes ($V_{PM}$,  $V^{quat}_{FM}$ and $V^{tern}_{FM}$) is not straightforward. The ideal PM state with randomly oriented atomic magnetic moments ($V_{PM}$) can not be achieved in experiment below the Curie temperature ($T_C$) due to the FM order. The lattice parameters for single crystal at this hypothetical demagnetized state below $T_C$ can be experimentally estimated by extrapolating the lattice parameters above $T_C$ via the Debye theory and the Gr\"{u}neisen relation \cite{ANDREEV199559,WASSERMAN1990237}. Similarly, a demagnetized FM state with FM domains pointing to all equivalent easy directions ($V^{quat}_{FM}$ or $V^{tern}_{FM}$) is not easy to obtain since the real experimental demagnetized states below $T_C$ exhibit different distribution of magnetic domains that depend on the demagnetization method \cite{Birss}. In the case of a model including only isotropic magnetic interactions, $\omega_s$ is also given by Eq.\ref{eq:ws000} for a single domain FM state without external magnetic field, see Section \ref{subsection:lmp}.

\subsection{Hexagonal (I)}

In Lacheisserie convention, the magnetoelastic energy of Hexagonal (I) crystals (point groups $6mm$, $622$, $\bar{6}2m$, $6/mmm$) reads\cite{booktremolet}
\begin{equation}
\begin{aligned}
   \frac{1}{V_0} E_{me}  & =  \frac{1}{3}b_1^{\alpha,0}(\epsilon_{xx}+\epsilon_{yy}+\epsilon_{zz})+\frac{\sqrt{2}}{3}b_2^{\alpha,0}(\epsilon_{zz}-\frac{1}{2}[\epsilon_{xx}+\epsilon_{yy}])\\
   & +\left[\frac{1}{\sqrt{2}}b_{1}^{\alpha,2}(\epsilon_{xx}+\epsilon_{yy}+\epsilon_{zz})+b_{2}^{\alpha,2}(\epsilon_{zz}-\frac{1}{2}[\epsilon_{xx}+\epsilon_{yy}])\right]\left(\alpha_z^2-\frac{1}{3}\right)\\
    & + b^{\epsilon,2}\left[\frac{1}{2}(\alpha_x^2-\alpha_y^2)(\epsilon_{xx}-\epsilon_{yy})+2\alpha_x\alpha_y\epsilon_{xy}\right]+2b^{\zeta,2}(\alpha_x\alpha_z\epsilon_{xz}+\alpha_y\alpha_z\epsilon_{yz}),
\label{eq:hex_I_universal}     
\end{aligned}
\end{equation}
where $b_1^{\alpha,0}$ and $b_2^{\alpha,0}$ are the isotropic magnetoelastic constants, and $b_{1}^{\alpha,2}$, $b_{2}^{\alpha,2}$, $b^{\epsilon,2}$ and $b^{\zeta,2}$ are the anisotropic magnetoelastic constants. As in the cubic symmetry case, within this notation the magnetoelastic energy can be decomposed  into isotropic and anisotropic contributions (Eq.\ref{eq:Eme_universal_2}). In this case, the isotropic part reads
\begin{equation}
    \begin{aligned}
      \frac{1}{V_0}  E_{me}^{iso} = \frac{1}{3}b_1^{\alpha,0}(\epsilon_{xx}+\epsilon_{yy}+\epsilon_{zz})+\frac{\sqrt{2}}{3}b_2^{\alpha,0}(\epsilon_{zz}-\frac{1}{2}[\epsilon_{xx}+\epsilon_{yy}]).
    \label{eq:Eme_hex_iso}
    \end{aligned}
\end{equation}
On the other hand, in Clark convention the magnetoelastic energy reads\cite{Clark}
\begin{equation}
\begin{aligned}
    \frac{1}{V_0} E_{me}  & =  b_{11}(\epsilon_{xx}+\epsilon_{yy})+b_{12}\epsilon_{zz}+b_{21}\left(\alpha_z^2-\frac{1}{3}\right)(\epsilon_{xx}+\epsilon_{yy})+b_{22}\left(\alpha_z^2-\frac{1}{3}\right)\epsilon_{zz}\\
    & + b_3\left[\frac{1}{2}(\alpha_x^2-\alpha_y^2)(\epsilon_{xx}-\epsilon_{yy})+2\alpha_x\alpha_y\epsilon_{xy}\right]+2b_4(\alpha_x\alpha_z\epsilon_{xz}+\alpha_y\alpha_z\epsilon_{yz}), 
\label{eq:E_me_hex_I}     
\end{aligned}
\end{equation}
where these magnetoelastic constants are related to those defined in Lacheisserie convention as
\begin{equation}
\begin{aligned}
    b_{11} & = \frac{1}{3} b_1^{\alpha,0}-\frac{1}{3\sqrt{2}} b_2^{\alpha,0},\\
    b_{12} & = \frac{1}{3} b_1^{\alpha,0}+\frac{\sqrt{2}}{3} b_2^{\alpha,0},\\
    b_{21} & = \frac{1}{\sqrt{2}} b_1^{\alpha,2}-\frac{1}{2} b_2^{\alpha,2},\\
    b_{22} & = \frac{1}{\sqrt{2}} b_1^{\alpha,2}+ b_2^{\alpha,2},\\
    b_{3} & =  b^{\epsilon,2},\quad b_4 =b^{\zeta,2}.
\label{eq:E_me_hex_I_2}     
\end{aligned}
\end{equation}
We see that $b_{11}$ and $b_{12}$ can be expressed as a linear combination of the isotropic magnetoelastic constants $b_1^{\alpha,0}$ and $b_2^{\alpha,0}$. This means that the Clark convention also fully decouples the isotropic and anisotropic magnetic interactions in the definition of the magnetoelastic constants. Therefore, we can rewrite the isotropic part of the magnetoelastic energy (Eq.\ref{eq:Eme_hex_iso}) as
\begin{equation}
    \begin{aligned}
      \frac{1}{V_0}  E_{me}^{iso} = b_{11}(\epsilon_{xx}+\epsilon_{yy})+b_{12}\epsilon_{zz}.
    \label{eq:Eme_hex_iso_2}
    \end{aligned}
\end{equation}
In MAELAS, we use this alternative expression in Eq.\ref{eq:E_tot} to derive the isotropic magnetoelastic constants $b_{11}$ and $b_{12}$, see Table \ref{tab:method_data}. Once these quantities are obtained, -mode 3 of MAELAS also calculates $b_1^{\alpha,0}$ and $b_2^{\alpha,0}$ by solving the system of equations given in Eq.\ref{eq:E_me_hex_I_2}. 

In Lacheisserie convention, the fractional change in length is written as\cite{booktremolet}
\begin{equation}
\begin{aligned}
     \frac{l- l_0}{l_0}\Bigg\vert_{\boldsymbol{\beta}}^{\boldsymbol{\alpha}} & =\frac{1}{3}\lambda_1^{\alpha,0}+\lambda_2^{\alpha,0}\left(\beta_z^2-\frac{1}{3}\right)+\lambda_1^{\alpha,2}\left(\alpha_z^2-\frac{1}{3}\right)\\
     & + \lambda_2^{\alpha,2}\left(\alpha_z^2-\frac{1}{3}\right)\left[\beta_z^2-\frac{1}{2}(\beta_x^2+\beta_y^2)\right]\\
     & +\lambda_{Lach}^{\epsilon,2}\left[\frac{1}{2}(\alpha_x^2-\alpha_y^2)(\beta_x^2-\beta_y^2)+2\alpha_x\alpha_y\beta_x\beta_y\right]\\
     & + 2\lambda^{\zeta,2}(\alpha_x\alpha_z\beta_x\beta_z+\alpha_y\alpha_z\beta_y\beta_z),
    \label{eq:delta_l_hex_I}
\end{aligned}
\end{equation}
where $\lambda_1^{\alpha,0}$ and $\lambda_2^{\alpha,0}$ are the isotropic magnetostrictive coefficients and $\lambda_1^{\alpha,2}$, $\lambda_2^{\alpha,2}$, $\lambda_{Lach}^{\epsilon,2}$ and $\lambda^{\zeta,2}$ are the anisotropic ones. On the other hand, the fractional change in length in Clark convention reads\cite{CLARK1980531,Clark}
\begin{equation}
\begin{aligned}
     \frac{l-l_0 }{l_0}\Bigg\vert_{\boldsymbol{\beta}}^{\boldsymbol{\alpha}} & =\lambda^{\alpha1,0}(\beta_x^2+\beta_y^2)+\lambda^{\alpha2,0}\beta_z^2+\lambda^{\alpha1,2}\left(\alpha_z^2-\frac{1}{3}\right)(\beta_x^2+\beta_y^2)\\
     & + \lambda^{\alpha2,2}\left(\alpha_z^2-\frac{1}{3}\right)\beta_z^2+\lambda^{\gamma,2}\left[\frac{1}{2}(\alpha_x^2-\alpha_y^2)(\beta_x^2-\beta_y^2)+2\alpha_x\alpha_y\beta_x\beta_y\right]\\
     & + 2\lambda_{Clark}^{\epsilon,2}(\alpha_x\alpha_z\beta_x\beta_z+\alpha_y\alpha_z\beta_y\beta_z),
    \label{eq:delta_l_hex_I_clark}
\end{aligned}
\end{equation}
where these magnetostrictive coefficients are related to those defined in Lacheisserie notation as \cite{CLARK1980531,booktremolet}
\begin{equation}
\begin{aligned}
     \lambda_1^{\alpha,0} & = 2\lambda^{\alpha1,0}+\lambda^{\alpha2,0},\\
     \lambda_2^{\alpha,0} & =  -\lambda^{\alpha1,0}+\lambda^{\alpha2,0},\\
    \lambda_1^{\alpha,2} & = 2\lambda^{\alpha1,2}+\lambda^{\alpha2,2},\\
    \lambda_2^{\alpha,2} & =  \frac{2}{3}(-\lambda^{\alpha1,2}+\lambda^{\alpha2,2}),\\
     \lambda_{Lach}^{\epsilon,2} & =  \lambda^{\gamma,2},\\
     \lambda^{\zeta,2} & = \lambda_{Clark}^{\epsilon,2}.
    \label{eq:lamb_Clark_hex_I}
\end{aligned}
\end{equation}
The isotropic magnetostrictive coefficients in Clark convention ($\lambda^{\alpha1,0}$ and $\lambda^{\alpha2,0}$) are associated to the isotropic magnetoelastic constants and elastic constants via\cite{CLARK1980531}
\begin{equation}
    \begin{aligned}
        \lambda^{\alpha1,0} & = \frac{-b_{11}C_{33}+b_{12}C_{13}}{C_{33}(C_{11}+C_{12})-2C_{13}^2},\\
        \lambda^{\alpha2,0} & = \frac{2b_{11}C_{13}-b_{12}(C_{11}+C_{12})}{C_{33}(C_{11}+C_{12})-2C_{13}^2}.
    \label{eq:lamb_hex_iso_correct}
    \end{aligned}
\end{equation}
In Clark's work\cite{CLARK1980531}, we note that the sign $-$ in front of the factor $b_{11}C_{33}$ in $\lambda^{\alpha1,0}$ is correctly written in the expression for the equilibrium strain tensor ($\epsilon_{xx}^{eq}$ and $\epsilon_{yy}^{eq}$), however was missing in the final expression for $\lambda^{\alpha1,0}$ incorrectly. Unfortunately, this misprint was further propagated to later publications like in Cullen et al.\cite{Cullen}, as well as in our publication of version 1.0 of MAELAS\cite{maelas_publication2021}. This also applies to other uniaxial crystal systems like Trigonal (I) and Tetragonal (I), where Eq.\ref{eq:lamb_hex_iso_correct} holds. We identified this problem by trying to verify the general formula Eq.\ref{eq:ws_maelas_gs} for uniaxial crystal systems since $\omega_s$ depends on $\lambda^{\alpha1,0}$, as shown below. 

\begin{figure}[h!]
\centering
\includegraphics[width=\columnwidth ,angle=0]{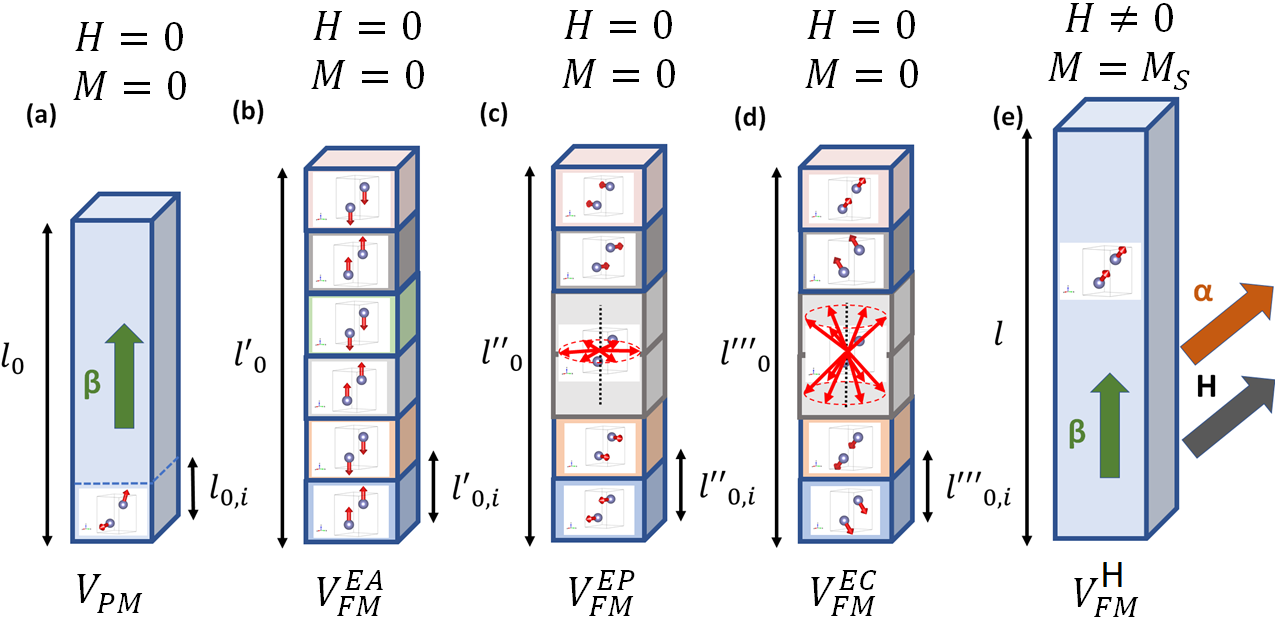}
\caption{(a) Ideal PM state of a hexagonal single crystal with randomly oriented atomic magnetic moments ($V_{PM}$). FM state with magnetic domains along all equivalent easy directions: (b) easy axis ($V_{FM}^{EA}$), (c) easy plane ($V_{FM}^{EP}$) and (d) easy cone ($V_{FM}^{EC}$). (e) Saturated magnetic state ($V_{FM}^{H}$). $M$, $M_s$ and $H$ represent the total magnetization, saturation magnetization and applied magnetic field, respectively.}
\label{fig:demag_hex}
\end{figure}

Applying a similar procedure to derive $\omega_s$ as in the cubic crystal (Eq.\ref{eq:ws00}) leads to the following result \cite{ANDREEV199559}
\begin{equation}
    \begin{aligned}
        \omega_s\simeq 2\lambda_a+\lambda_c,
    \label{eq:ws_hex}
    \end{aligned}
\end{equation}
where $\lambda_c$ is the  fractional change in length along the lattice vector $\boldsymbol{c}$, that is, $\boldsymbol{\beta}=(0,0,1)$. Again, $\omega_s$ depends on the definition of FM state. For instance, if we consider a single crystal with easy cone (cone angle $\Omega$ with respect to $\boldsymbol{c}$) where  FM domains point to all equivalent easy directions (demagnetized state), then we have \cite{maelas_v2,ANDREEV199559}
\begin{equation}
\begin{aligned}
     \lambda_a & = \frac{l'''_0-l_0}{l_0}\Bigg\vert_{\boldsymbol{\beta}=(1,0,0)} =\lambda^{\alpha1,0} +\left(\cos^2\Omega-\frac{1}{3}\right)\lambda^{\alpha1,2},\\
     \lambda_c & = \frac{l'''_0-l_0}{l_0}\Bigg\vert_{\boldsymbol{\beta}=(0,0,1)} =\lambda^{\alpha2,0}+\left(\cos^2\Omega-\frac{1}{3}\right)\lambda^{\alpha2,2},
    \label{eq:sum_delta_l_hex_5}
\end{aligned}
\end{equation}
where $l'''_0$ is the length at this considered FM state (volume $V^{EC}_{FM}$), and $l_0$ is the length at the PM-like state (randomly oriented atomic magnetic moments). Therefore, replacing Eq.\ref{eq:sum_delta_l_hex_5} in Eq.\ref{eq:ws_hex} leads to\cite{ANDREEV199559}
\begin{equation}
    \begin{aligned}
       \frac{V^{EC}_{FM}-V_{PM}}{V_{PM}}= \omega_s \simeq \omega_s^{iso}+\omega_s^{ani},
    \label{eq:ws_hex_total}
    \end{aligned}
\end{equation}
where 
\begin{equation}
    \begin{aligned}
        \omega_s^{iso} & = 2\lambda^{\alpha1,0}+\lambda^{\alpha2,0},\\
        \omega_s^{ani} & = 2\left(\cos^2\Omega-\frac{1}{3}\right)\lambda^{\alpha1,2}+\left(\cos^2\Omega-\frac{1}{3}\right)\lambda^{\alpha2,2}.
    \label{eq:ws_hex_iso_ani}
    \end{aligned}
\end{equation}
Note that the cases with easy axis and easy plane are obtained by setting the cone angle $\Omega=0$ and 	 $\Omega=\pi/2$ in Eq.\ref{eq:ws_hex_iso_ani}, respectively. For example, these cases were experimentally analyzed by Andreev for Rare-Earth intermetallics with Co and Fe \cite{ANDREEV199559}. These magnetic states are schematically represented in Fig.\ref{fig:demag_hex}. The quantity $\omega_s^{iso}$ comes from the isotropic magnetic interactions and is independent of the FM state and easy directions. The -mode 3 of MAELAS computes $\omega_s^{iso}$ from the calculated isotropic magnetostrictive coefficients ($\lambda^{\alpha1,0}$ and $\lambda^{\alpha2,0}$) via Eq.\ref{eq:ws_hex_iso_ani} automatically. Similarly, the anisotropic contribution $\omega_s^{ani}$ is automatically  calculated through -mode 1 or -mode 2 for the easy axis and easy plane cases.

\subsection{Trigonal (I) and Tetragonal (I)}

The analysis of the isotropic contribution to the magnetoelastic energy for Trigonal (I) and Tetragonal (I) systems is the same as in the Hexagonal (I) symmetry\cite{Cullen,booktremolet}. For these systems $E_{me}^{iso}$ and $\omega_s$ are also given by Eqs.\ref{eq:Eme_hex_iso_2} and \ref{eq:ws_hex_total}, respectively. Similarly, the correct expression of $\lambda^{\alpha1,0}$ for these crystal systems is given by Eq.\ref{eq:lamb_hex_iso_correct}, which was incorrectly written in previous publications\cite{Cullen,maelas_publication2021}.

\subsection{Orthorhombic}

In Lacheisserie convention, the magnetoelastic energy of Orthorhombic crystals is given by\cite{booktremolet}
\begin{equation}
\begin{aligned}
    \frac{1}{V_0} E_{me}  & =  \frac{1}{V_0}(E^{iso}_{me}+E^{ani}_{me})
\label{eq:E_me_orth}     
\end{aligned}
\end{equation}
where the isotropic term is
\begin{equation}
\begin{aligned}
     E^{iso}_{me}  & =  \frac{1}{3}b_1^{\alpha,0}(\epsilon_{xx}+\epsilon_{yy}+\epsilon_{zz})+\frac{\sqrt{2}}{3}b_2^{\alpha,0}\left(\epsilon_{zz}-\frac{1}{2}[\epsilon_{xx}+\epsilon_{yy}]\right)+\\& + \frac{1}{\sqrt{6}}b_3^{\alpha,0}\left(\epsilon_{xx}-\epsilon_{yy}\right),
\label{eq:E_me_orth_iso}     
\end{aligned}
\end{equation}
and the anisotropic term is
\begin{equation}
\begin{aligned}
     E^{ani}_{me}  & =  [\frac{\sqrt{2}}{3}b_1^{\alpha,2}(\epsilon_{xx}+\epsilon_{yy}+\epsilon_{zz})+\frac{2}{3}b_2^{\alpha,2}\left(\epsilon_{zz}-\frac{1}{2}[\epsilon_{xx}+\epsilon_{yy}]\right)+\\& + \frac{1}{\sqrt{6}}b_3^{\alpha,2}\left(\epsilon_{xx}-\epsilon_{yy}\right)]\left[\alpha_z^2-\frac{1}{2}(\alpha_{x}^2+\alpha_y^2)\right]\\
     & + [\frac{1}{\sqrt{6}}b_1^{\alpha,2'}(\epsilon_{xx}+\epsilon_{yy}+\epsilon_{zz})+\frac{1}{\sqrt{3}}b_2^{\alpha,2'}\left(\epsilon_{zz}-\frac{1}{2}[\epsilon_{xx}+\epsilon_{yy}]\right)+\\ 
     & + \frac{1}{2}b_3^{\alpha,2'}\left(\epsilon_{xx}-\epsilon_{yy}\right)]\left[\alpha_{x}^2-\alpha_y^2\right]\\
     & + 2b^{\beta,2}\alpha_x\alpha_y\epsilon_{xy}+2b^{\delta,2}\alpha_x\alpha_z\epsilon_{xz}+2b^{\gamma,2}\alpha_y\alpha_z\epsilon_{yz}.
\label{eq:E_me_orth_ani}     
\end{aligned}
\end{equation}
On the other hand, the magnetoelastic energy defined in the version 1.0 of MAELAS (Nieves convention) is given by\cite{maelas_publication2021}
\begin{equation}
\begin{aligned}
    \frac{1}{V_0}E_{me}  & =  b_{01}\epsilon_{xx}+b_{02}\epsilon_{yy}+b_{03}\epsilon_{zz}+b_1\alpha_x^2\epsilon_{xx}+b_2\alpha_y^2\epsilon_{xx}+b_3\alpha_x^2\epsilon_{yy}+b_4\alpha_y^2\epsilon_{yy}\\
    & +b_5\alpha_x^2\epsilon_{zz}+b_6\alpha_y^2\epsilon_{zz}+2b_7\alpha_x\alpha_y\epsilon_{xy}+2b_8\alpha_x\alpha_z\epsilon_{xz}+2b_9\alpha_y\alpha_z\epsilon_{yz}.
\label{eq:E_me_ortho_nieves}     
\end{aligned}
\end{equation}
These magnetoelastic constants are related to those defined in Eq.\ref{eq:E_me_orth} thought the following equations
\begin{equation}
\begin{aligned}
b_{01} & = \frac{1}{3}(b_1^{\alpha,0} + \sqrt{2} b_1^{\alpha,2}-\frac{1}{\sqrt{2}} b_2^{\alpha,0} - b_2^{\alpha,2}) + \frac{1}{\sqrt{6}} (b_3^{\alpha,0} + b_3^{\alpha,2}),\\
b_{02} & = \frac{1}{3}(b_1^{\alpha,0} + \sqrt{2} b_1^{\alpha,2}-\frac{1}{\sqrt{2}} b_2^{\alpha,0} - b_2^{\alpha,2}) - \frac{1}{\sqrt{6}} (b_3^{\alpha,0} + b_3^{\alpha,2}),\\
b_{03} & = \frac{1}{3}(b_1^{\alpha,0} + \sqrt{2} b_1^{\alpha,2}+\sqrt{2} b_2^{\alpha,0} + 2b_2^{\alpha,2}),\\
b_1 & = \frac{1}{6} (3 b_3^{\alpha,2'} + 3 b_2^{\alpha,2} - \sqrt{3} b_2^{\alpha,2'} + \sqrt{6} b_1^{\alpha,2'} - \frac{3}{\sqrt{2}}[\sqrt{3} b_3^{\alpha,2} + 2 b_1^{\alpha,2}  ]),\\
b_2 & = \frac{1}{2} (-\sqrt{\frac{3}{2}}b_3^{\alpha,2}  - b_3^{\alpha,2'} + b_2^{\alpha,2}  + \frac{1}{\sqrt{3}}b_2^{\alpha,2'}  - 
     \sqrt{2}b_1^{\alpha,2}  - \sqrt{\frac{2}{3}}b_1^{\alpha,2'} ),\\
b_3 & = \frac{1}{6} (-3 b_3^{\alpha,2'} + 3 b_2^{\alpha,2} - \sqrt{3} b_2^{\alpha,2'} + \sqrt{6} b_1^{\alpha,2'} - \frac{3}{\sqrt{2}}[-\sqrt{3} b_3^{\alpha,2} + 2 b_1^{\alpha,2}  ]),\\
b_4 & = \frac{1}{2} (\sqrt{\frac{3}{2}}b_3^{\alpha,2}  + b_3^{\alpha,2'} + b_2^{\alpha,2}  + \frac{1}{\sqrt{3}}b_2^{\alpha,2'}  - 
     \sqrt{2}b_1^{\alpha,2}  - \sqrt{\frac{2}{3}}b_1^{\alpha,2'} ),\\
b_5 & = \frac{1}{6}  (-6 b_2^{\alpha,2}  + 2 \sqrt{3} b_2^{\alpha,2'} + \sqrt{6}b_1^{\alpha,2'} - 3\sqrt{2} b_1^{\alpha,2}  ),\\
b_6 & = \frac{1}{2}  (-2 b_2^{\alpha,2}  - \frac{2}{\sqrt{3}} b_2^{\alpha,2'} - \sqrt{2}b_1^{\alpha,2} - \sqrt{\frac{2}{3}} b_1^{\alpha,2'}  ),\\
b_7 & = b^{\beta,2},\quad b_8  = b^{\delta,2},\quad b_9  = b^{\gamma,2},
\label{eq:E_me_orth_nieves_lach}     
\end{aligned}
\end{equation}
which are found by changing $\alpha_z^2$  in Eq.\ref{eq:E_me_orth_ani} to $1-\alpha_x^2-\alpha_y^2$, and collecting the terms in Eq.\ref{eq:E_me_orth} in the same way as in Eq.\ref{eq:E_me_ortho_nieves}. We see that $b_{01}$, $b_{02}$ and $b_{03}$ are expressed as a linear combination of isotropic ($b_i^{\alpha,0}$) and anisotropic ($b_i^{\alpha,2}$) magnetoelastic constants in Lacheisserie convention. This means that the definition of the isotropic magnetoelastic constants in Nieves convention ($b_{01}$, $b_{02}$ and $b_{03}$) mixes the isotropic and anisotropic magnetic interactions. Consequently, -mode 3 of MAELAS uses the Lacheisserie isotropic magnetoelastic energy ($E_{me}^{iso}$) given by Eq.\ref{eq:E_me_orth_iso} in Eq.\ref{eq:E_tot} to determine the isotropic magnetoelastic constants ($b_i^{\alpha,0}$, $i=1,2,3$) automatically, see Tables \ref{tab:method_data} and \ref{tab:calculation_data}. The magnetoelastic constants $b_{01}$, $b_{02}$ and $b_{03}$ can be manually calculated by inserting the obtained isotropic and anisotropic magnetoelastic constants in Lacheisserie convention (the isotropic ones via -mode 3 and the anisotropic ones via -mode 1 or -mode 2) into Eq.\ref{eq:E_me_orth_nieves_lach}, as shown in Table \ref{tab:manual_data}.

In Lacheisserie convention, the fractional change in length  reads\cite{booktremolet}
\begin{equation}
\begin{aligned}
     \frac{l- l_0}{l_0}\Bigg\vert_{\boldsymbol{\beta}}^{\boldsymbol{\alpha}} & =\frac{1}{3}\lambda_1^{\alpha,0}+\lambda_2^{\alpha,0}\left(\beta_z^2-\frac{1}{3}\right)+\lambda_3^{\alpha,0}(\beta_x^2-\beta_y^2)\\
     & +\left[\frac{1}{3}\lambda_1^{\alpha,2}+\lambda_2^{\alpha,2}\left(\beta_z^2-\frac{1}{3}\right)+\lambda_3^{\alpha,2}(\beta_x^2-\beta_y^2)\right]\left[\alpha_z^2-\frac{1}{2}(\alpha_{x}^2+\alpha_y^2)\right]\\
     & +\left[\frac{1}{3}\lambda_1^{\alpha,2'}+\lambda_2^{\alpha,2'}\left(\beta_z^2-\frac{1}{3}\right)+\lambda_3^{\alpha,2'}(\beta_x^2-\beta_y^2)\right]\left[\frac{1}{2}(\alpha_{x}^2-\alpha_y^2)\right]\\
     & +2\lambda^{\beta,2}\alpha_x\alpha_y\beta_x\beta_y +2 \lambda^{\delta,2}\alpha_x\alpha_z\beta_x\beta_z+2\lambda^{\gamma,2}\alpha_y\alpha_z\beta_y\beta_z,
    \label{eq:delta_l_ortho}
\end{aligned}
\end{equation}
while in Mason convention is given by\cite{Mason,maelas_publication2021}
\begin{equation}
\begin{aligned}
     \frac{l- l_0}{l_0}\Bigg\vert_{\boldsymbol{\beta}}^{\boldsymbol{\alpha}} & =\lambda^{\alpha1,0}\beta_x^2+\lambda^{\alpha2,0}\beta_y^2+\lambda^{\alpha3,0}\beta_z^2+\lambda_1(\alpha_x^2\beta_x^2-\alpha_x\alpha_y\beta_x\beta_y-\alpha_x\alpha_z\beta_x\beta_z)\\
     & +\lambda_2(\alpha_y^2\beta_x^2-\alpha_x\alpha_y\beta_x\beta_y)+\lambda_3(\alpha_x^2\beta_y^2-\alpha_x\alpha_y\beta_x\beta_y)\\
     & +\lambda_4(\alpha_y^2\beta_y^2-\alpha_x\alpha_y\beta_x\beta_y-\alpha_y\alpha_z\beta_y\beta_z)+\lambda_5(\alpha_x^2\beta_z^2-\alpha_x\alpha_z\beta_x\beta_z)\\
     & + \lambda_6(\alpha_y^2\beta_z^2-\alpha_y\alpha_z\beta_y\beta_z)+4\lambda_7\alpha_x\alpha_y\beta_x\beta_y+4\lambda_8\alpha_x\alpha_z\beta_x\beta_z+4\lambda_9\alpha_y\alpha_z\beta_y\beta_z.
    \label{eq:delta_l_ortho_mason}
\end{aligned}
\end{equation}
The relationships between these magnetostrictive coefficients are  
\begin{equation}
\begin{aligned}
\lambda^{\alpha1,0} & = \frac{1}{3} (\lambda_1^{\alpha,0} + \lambda_1^{\alpha,2} - \lambda_2^{\alpha,0} - \lambda_2^{\alpha,2}) + \lambda_3^{\alpha,0} + \lambda_3^{\alpha,2},\\
\lambda^{\alpha2,0} & = \frac{1}{3} (\lambda_1^{\alpha,0} + \lambda_1^{\alpha,2} - \lambda_2^{\alpha,0} - \lambda_2^{\alpha,2}) - \lambda_3^{\alpha,0} - \lambda_3^{\alpha,2},\\
\lambda^{\alpha3,0} & = \frac{1}{3} (\lambda_1^{\alpha,0} + \lambda_1^{\alpha,2} + 2\lambda_2^{\alpha,0} +2 \lambda_2^{\alpha,2}),\\
\lambda_1 & = \frac{1}{6} (-3 \lambda_1^{\alpha,2} + \lambda_1^{\alpha,2'}  + 3 \lambda_2^{\alpha,2}  - \lambda_2^{\alpha,2'}  - 9 \lambda_3^{\alpha,2}  + 
     3 \lambda_3^{\alpha,2'}),\\
 \lambda_2 & = -\frac{1}{6} (3 \lambda_1^{\alpha,2} + \lambda_1^{\alpha,2'}  - 3 \lambda_2^{\alpha,2}  - \lambda_2^{\alpha,2'}  + 9 \lambda_3^{\alpha,2}  + 
     3 \lambda_3^{\alpha,2'}),\\  
\lambda_3 & = \frac{1}{6} (-3 \lambda_1^{\alpha,2} + \lambda_1^{\alpha,2'}  + 3 \lambda_2^{\alpha,2}  - \lambda_2^{\alpha,2'}  + 9 \lambda_3^{\alpha,2}  - 
     3 \lambda_3^{\alpha,2'}),\\
\lambda_4 & = -\frac{1}{6} (3 \lambda_1^{\alpha,2} + \lambda_1^{\alpha,2'}  - 3 \lambda_2^{\alpha,2}  - \lambda_2^{\alpha,2'}  - 9 \lambda_3^{\alpha,2}  - 
     3 \lambda_3^{\alpha,2'}),\\ 
\lambda_5 & = \frac{1}{6} (-3 \lambda_1^{\alpha,2} + \lambda_1^{\alpha,2'} - 6 \lambda_2^{\alpha,2} + 2 \lambda_2^{\alpha,2'} ),\\
\lambda_6 & = -\frac{1}{6} (3 \lambda_1^{\alpha,2} + \lambda_1^{\alpha,2'} + 6 \lambda_2^{\alpha,2} + 2 \lambda_2^{\alpha,2'} ),\\
\lambda_7 & = \frac{1}{2}(\lambda^{\beta,2} -\lambda_1^{\alpha,2}+\lambda_2^{\alpha,2}),\\
\lambda_8 & = \frac{1}{4}(2\lambda^{\gamma,2} + \lambda_4 + \lambda_6),\\
\lambda_9 & = \frac{1}{4}(2\lambda^{\delta,2} + \lambda_1 + \lambda_5).
\label{eq:lmb_mason_lach}     
\end{aligned}
\end{equation}
If the elastic tensor is provided \cite{AELAS}, then -mode 3 of MAELAS  determines the Lacheisserie isotropic magnetostrictive coefficients ($\lambda_i^{\alpha,0}$, $i=1,2,3$) automatically (see Table \ref{tab:calculation_data}) through the relationships between the isotropic magnetoelastic constants ($b_i^{\alpha,0}$, $i=1,2,3$) and elastic constants ($C_{ij}$) given in Ref.\cite{booktremolet}, see Fig.\ref{fig:diagram_3_modes}. The magnetostrictive coefficients  $\lambda^{\alpha i,0}$, $i=1,2,3$ can be manually calculated by inserting the obtained isotropic and anisotropic magnetostrictive coefficients in Lacheisserie convention (the isotropic ones via -mode 3 and the anisotropic ones via -mode 1 or -mode 2) into Eq.\ref{eq:lmb_mason_lach}, as shown in Table \ref{tab:manual_data}. 

\begin{figure}[h!]
\centering
\includegraphics[width=\columnwidth ,angle=0]{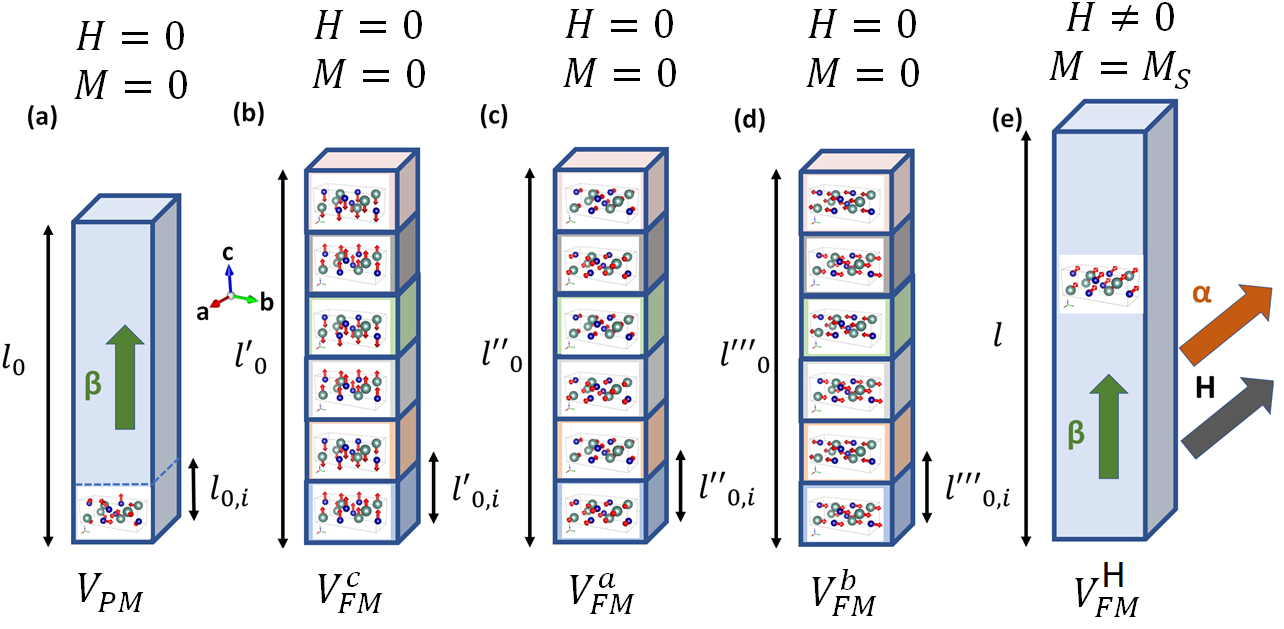}
\caption{(a) Ideal PM state of a orthorhombic single crystal with randomly oriented atomic magnetic moments ($V_{PM}$). FM state with magnetic domains along all equivalent easy directions: (b) easy axis along the lattice vector $\boldsymbol{c}$ parallel to z-axis ($V_{FM}^{c}$), (c) easy axis along the lattice vector $\boldsymbol{a}$ parallel to x-axis ($V_{FM}^{a}$) and (d) easy axis along the lattice vector $\boldsymbol{b}$ parallel to y-axis ($V_{FM}^{b}$).  (e) Saturated magnetic state ($V_{FM}^{H}$). MAELAS uses the IEEE lattice convention $c<a<b$.}
\label{fig:demag_orto}
\end{figure}

Applying a similar procedure to derive $\omega_s$ as in the cubic crystal (Eq.\ref{eq:ws00}) leads to the following result \cite{ANDREEV199559}
\begin{equation}
    \begin{aligned}
        \omega_s\simeq \lambda_a+\lambda_b+\lambda_c,
    \label{eq:ws_orto}
    \end{aligned}
\end{equation}
where $\lambda_b$ is the  fractional change in length along the lattice vector $\boldsymbol{b}$, that is, $\boldsymbol{\beta}=(0,1,0)$. We assume the same IEEE lattice convention ($c<a<b$) that is used in MAELAS and AELAS \cite{AELAS,maelas_publication2021,maelas_v2}. Again, $\omega_s$ depends on the definition of FM state. For instance, if we consider a FM state with easy axis along the lattice vector $\boldsymbol{c}=(0,0,c)$ where  FM domains point to all equivalent easy directions $\boldsymbol{\alpha}=(0,0,\pm1)$ (demagnetized state), then we have \cite{maelas_v2}
\begin{equation}
\begin{aligned}
     \lambda_a & = \frac{l'_0-l_0}{l_0}\Bigg\vert_{\boldsymbol{\beta}=(1,0,0)} =\frac{1}{3}\lambda_1^{\alpha,0}-\frac{1}{3}\lambda_2^{\alpha,0}+\lambda_3^{\alpha,0}+\frac{1}{3}\lambda_1^{\alpha,2}-\frac{1}{3}\lambda_2^{\alpha,2}+\lambda_3^{\alpha,2},\\
     \lambda_b & = \frac{l'_0-l_0}{l_0}\Bigg\vert_{\boldsymbol{\beta}=(0,1,0)} =\frac{1}{3}\lambda_1^{\alpha,0}-\frac{1}{3}\lambda_2^{\alpha,0}-\lambda_3^{\alpha,0}+\frac{1}{3}\lambda_1^{\alpha,2}-\frac{1}{3}\lambda_2^{\alpha,2}-\lambda_3^{\alpha,2},\\
     \lambda_c & = \frac{l'_0-l_0}{l_0}\Bigg\vert_{\boldsymbol{\beta}=(0,0,1)} =\frac{1}{3}\lambda_1^{\alpha,0}+\frac{2}{3}\lambda_2^{\alpha,0}+\frac{1}{3}\lambda_1^{\alpha,2}+\frac{2}{3}\lambda_2^{\alpha,2},
    \label{eq:sum_delta_l_orto}
\end{aligned}
\end{equation}
where $l'_0$ is the length at this considered FM state (volume $V^{c}_{FM}$), and $l_0$ is the length at the PM-like state (randomly oriented atomic magnetic moments). Therefore, replacing  Eq.\ref{eq:sum_delta_l_orto} in Eq.\ref{eq:ws_orto} leads to
\begin{equation}
    \begin{aligned}
       \frac{V^{c}_{FM}-V_{PM}}{V_{PM}}= \omega_s \simeq \omega_s^{iso}+\omega_s^{ani},
    \label{eq:ws_orto_total}
    \end{aligned}
\end{equation}
where 
\begin{equation}
    \begin{aligned}
        \omega_s^{iso} & = \lambda_1^{\alpha,0},\\
        \omega_s^{ani} & =  \lambda_1^{\alpha,2}.
    \label{eq:ws_orto_iso_ani_c}
    \end{aligned}
\end{equation}
Similarly, if we consider a FM state with easy axis along the lattice vector $\boldsymbol{a}=(a,0,0)$  where  FM domains point to all equivalent easy directions $\boldsymbol{\alpha}=(\pm1,0,0)$ (demagnetized state, volume $V^{a}_{FM}$), then we find
\begin{equation}
    \begin{aligned}
        \frac{V^{a}_{FM}-V_{PM}}{V_{PM}} & = \omega_s \simeq \omega_s^{iso}+\omega_s^{ani},\\
        \omega_s^{iso} & = \lambda_1^{\alpha,0},\\
        \omega_s^{ani} & =  -\frac{1}{2}\lambda_1^{\alpha,2}+\frac{1}{2}\lambda_1^{\alpha,2'}.
    \label{eq:ws_orto_iso_ani_a}
    \end{aligned}
\end{equation}
Lastly, if we consider a FM state with easy axis along the lattice vector $\boldsymbol{b}=(0,b,0)$  where  FM domains point to all equivalent easy directions $\boldsymbol{\alpha}=(0,\pm1,0)$ (demagnetized state, volume $V^{b}_{FM}$), then we find
\begin{equation}
    \begin{aligned}
        \frac{V^{b}_{FM}-V_{PM}}{V_{PM}} & = \omega_s \simeq \omega_s^{iso}+\omega_s^{ani},\\
        \omega_s^{iso} & = \lambda_1^{\alpha,0},\\
        \omega_s^{ani} & =  -\frac{1}{2}\lambda_1^{\alpha,2}-\frac{1}{2}\lambda_1^{\alpha,2'}.
    \label{eq:ws_orto_iso_ani_b}
    \end{aligned}
\end{equation}
These magnetic states are schematically represented in Fig.\ref{fig:demag_orto}.  The -mode 3 of MAELAS computes $\omega_s^{iso}$ from the calculated isotropic magnetostrictive coefficients automatically, while the anisotropic contribution $\omega_s^{ani}$ is automatically  calculated through -mode 1 or -mode 2 for these three cases.

\bibliographystyle{elsarticle-num}
\bibliography{mybibfile.bib}







\end{document}